\documentclass{aastex61}
\usepackage{amsmath,amsthm,amssymb}
\usepackage{titlesec}
\usepackage{graphicx}
\usepackage{tabularx}
\usepackage{float}
\usepackage{color}
\usepackage{scalefnt}
\usepackage{hyperref}
\usepackage[numbib]{tocbibind}
\usepackage{enumitem}
\usepackage{multirow}
\usepackage{mathrsfs}
\usepackage{wasysym}

\begin{document}

\title{Exo-Milankovitch Cycles I: Orbits and Rotation States}
\shorttitle{Exo-Milankovitch I}
\author{Russell Deitrick}
\affil{Department of Astronomy, University of Washington, Seattle, WA 98195-1580, USA}
\affil{Virtual Planetary Laboratory, University of Washington, Seattle, WA 98195-1580, USA}
\email{deitrr@astro.washington.edu}
\author{Rory Barnes}
\affil{Department of Astronomy, University of Washington, Seattle, WA 98195-1580, USA}
\affil{Virtual Planetary Laboratory, University of Washington, Seattle, WA 98195-1580, USA}
\author{Thomas R. Quinn}
\affil{Department of Astronomy, University of Washington, Seattle, WA 98195-1580, USA}
\affil{Virtual Planetary Laboratory, University of Washington, Seattle, WA 98195-1580, USA}
\author{John Armstrong}
\affil{Department of Physics, Weber State University, Ogden, UT 84408-2508, USA}
\affil{Virtual Planetary Laboratory, University of Washington, Seattle, WA 98195-1580, USA}
\author{Benjamin Charnay}
\affil{LESIA, Observatoire de Paris, PSL Research University, CNRS, Sorbonne Universit\'{e}s, UPMC Univ. Paris 06, Univ. Paris Diderot, Sorbonne, Paris Cit\'{e}, 5 Place Jules Janssen, 92195 Meudon, France}
\affil{Virtual Planetary Laboratory, University of Washington, Seattle, WA 98195-1580, USA}
\author{Caitlyn Wilhelm}
\affil{Department of Astronomy, University of Washington, Seattle, WA 98195-1580, USA}
\affil{Virtual Planetary Laboratory, University of Washington, Seattle, WA 98195-1580, USA}
\date{}

\begin{abstract}
The obliquity of the Earth, which controls our seasons, varies by only $\sim$2.5$^{\circ}$ over $\sim$40,000 years, and its eccentricity varies by only $\sim 0.05$ over $100,000$ years. Nonetheless, these small variations influence Earth's ice ages. For exoplanets, however, variations can be significantly larger. Previous studies of the habitability of moonless Earth-like exoplanets have found that high obliquities, high eccentricities, and dynamical variations can extend the outer edge of the habitable zone by preventing runaway glaciation (snowball states). We expand upon these studies by exploring the orbital dynamics with a semi-analytic model that allows us to map broad regions of parameter space. We find that in general, the largest drivers of obliquity variations are secular spin-orbit resonances. We show how the obliquity varies in several test cases, including Kepler-62 f, across a wide range of orbital and spin parameters. These obliquity variations, alongside orbital variations, will have a dramatic impact on the climates of such planets. 
\end{abstract}

\maketitle

\section{INTRODUCTION}
The habitable zone (HZ) is the region around a star in which a planet with a roughly Earth-like atmosphere (i.e., dominated by N$_2$, CO$_2$, and water) can maintain liquid water on its surface for a significant billions of years \citep{huang1959}. Though initially this concept was discussed only for the sun and sun-like stars, it was expanded to include stellar types F-M by \cite{kasting1993}. Their study used a one-dimensional radiation-convective model to simulate the climates of Earth-like planets orbiting stars of types F, G, K, and M at different orbital distances to determine where the surface temperature was in the correct range for stable liquid water. This definition of the HZ does not include the possibility of subsurface life, such as life that might exist on Mars or Europa; however, currently this definition is appropriate for exoplanets because it is likely to be much more difficult to detect \emph{subsurface} life across interstellar distances. 
Many studies have since used one-dimensional radiative-convective models to expand on and refine this work \citep{selsis2007, kopparapu2013, goldblatt2013, kopparapu2014}. 

These HZ definitions are based on the average stellar flux received by a planet, and do not fully account for the effects of eccentricity and obliquity. To understand the importance of these parameters, it is worth reviewing the research done on Earth's ice ages and similar theories applied to exoplanets.

The connection between Earth's ice ages and its orbital/obliquity evolution was first suggested by Joseph Alphonse Adh\'{e}mar in 1842. This theory was continued by James Croll in the second half of the nineteenth century, but it wasn't until the twentieth century, when Milutin Milankovi\'{c} performed the first highly accurate calculations of Earth's insolation history and subsequently collaborated with climate scientist Wladimir K\"oppen, that the theory achieved much success. Hence the theory became commonly known as ``Milankovitch theory'' and the insolation variations were dubbed ``Milankovitch cycles''.

Milankovi\'c and K\"oppen identified what is still believed to be the primary connection between orbital variations and the ice ages: summer-time insolation at high latitude. During summer, if high latitude insolation is strong enough to raise temperatures high enough, for a long enough period of time, to melt all snow and ice, then from year to year, ice sheets on land will shrink. Conversely, if the insolation is weaker and temperatures are cooler in summer, then ice will tend to grow from year to year, leading to larger ice sheets.

After Milankovi\'c, calculations of Earth's orbital and insolation history continued to improve \citep[some examples are][]{brouwer1950,sharaf1967,bretagnon1974,berger1978,berger1991}. The development of the dynamical theory of the Earth-Moon-Sun system by \cite{kino1975,kino1977} facilitated  perhaps the most accurate determination of the Earth's orbital and obliquity evolution by \cite{laskar1986} and \cite{laskar1993}.

There is, however, a great deal of controversy surrounding Milankovitch theory \citep[see][]{wunsch2004, maslin2016}. While it is clear that eccentricity, obliquity, and precession can affect Earth's climate, it is unclear how the rather small variations in these parameters (for Earth) trigger or influence such dramatic changes. Several theories have proven tentatively successful at explaining this \citep{clark1998,roe2006,tziperman2006,huybers2008}, but a detailed review is beyond the scope of this work. For now, we will move on to exoplanets, which can have much more dramatic orbital variations than Earth.

Habitable exoplanets may not, for example, have a large satellite like Earth's Moon. The importance of the Moon on Earth's obliquity was pointed out by \cite{ward1974}. \cite{laskar1993,laskar1993b} (and later, \cite{lissauer2012}) showed that, without the Moon, Earth's obliquity could oscillate by a much larger amount ($\sim 20^{\circ}$ over $\sim 1$ Myr, rather than the $\sim 2.5^{\circ}$ we find with the Moon). As the small oscillation Earth experiences over $\sim 40$ kyr seems to influence the ice ages, it was expected, by extrapolation, that this much larger oscillation would be catastrophic. 

Habitable exoplanets should not be expected to have large moons like Earth's, as we do not yet understand the probability of the Moon-forming impact. Thus it is important to try to quantify the effects of obliquities very different from Earth's. 

 One of the first studies to explore the effect of obliquity on habitability was \cite{williamskasting97}. These authors used a 1D EBM with a parameterized carbon cycle to account for effects obliquity can have on weathering rates and thus changes in atmospheric CO$_2$ pressures. This paper addressed the problem that was proposed by \cite{laskar1993b}, that Earth might be uninhabitable without the Moon to stabilize its obliquity. \cite{williamskasting97} concluded that Earth could remain habitable with a wide range of obliquities ($23.5^{\circ}-90^{\circ}$), but suggested that large regions at the poles would be frozen for extended periods of time at obliquity $=90^{\circ}$. 

\cite{williams2003} furthered this study of the obliquity effects on climate using a 3D GCM. Though they did not model a large range of solar fluxes, they did find a number of interesting results. Like \cite{williamskasting97}, they found that an Earth-like planet at the current solar flux can remain habitable with any obliquity---none of their simulations experienced a snowball or runaway greenhouse event. Additionally, they noted that at high obliquity, the planet can develop ice-belts, i.e. ice sheets and sea ice at the equator, rather than ice caps at the poles. In fact, there is some evidence indicating the formation of tropical and mid-latitude glaciers on Mars \citep{head2006,forget2006}, which may have occurred during high obliquity times.

Further studies have investigated the climate response to different obliquities. \cite{spiegel2009} used a 1D EBM to find the boundaries of the HZ across the range of obliquity $0^{\circ}-90^{\circ}$. They found that the edges of the HZ can change by as much as $10\%$, depending on the obliquity and the efficiency of heat diffusion, and showed that  high obliquity planets have high latitudes which are warmer on average than the equatorial regions, but experience extreme seasonal variations. \cite{rose2017} examined glaciation under different obliquities using an analytical EBM, showing that equatorial ice is much less stable and less likely than polar ice caps. 

To lowest order, the effect of eccentricity on climate can be understood by examining the annually-averaged stellar flux. The mean stellar flux goes as $S \propto (1-e^2)^{-1/2}$, where $e$ is the eccentricity \citep{laskar1993}. Despite the fact that a planet spends more time at apocenter than at pericenter (Kepler's second law), increasing the orbital eccentricity actually \emph{increases} the amount of flux the planet receives over an orbit. In reality, the effect of eccentricity on climate is more complicated than this simple picture, because of seasonal variations. For example, the duration and intensity of seasons will depend on the shape of the planet's orbit, potentially leading to nonlinear effects on the climate. Further, while the average stellar flux is informative, the primary quantity of interest is temperature: we would like to know whether liquid water can be maintained on the surface of the planet. One informative value is the equilibrium temperature---the temperature of the pressure level at which the atmosphere becomes optically thick. As noted by \cite{selsis2007}, above $T_{eq} \approx 270$ K, a planet with an Earth-like atmosphere would enter a runaway greenhouse. However, the equilibrium temperature can sometimes be deceptive; for example, Venus's $T_{eq} \approx 227$ K. Rather, the desired quantity is the surface temperature of the planet, and to estimate that, a climate model is necessary, as the relationship with the incoming flux and surface temperature is extremely complicated because of the greenhouse effect, atmospheric circulation, precipitation, etc. We turn now to several studies that have quantified the relationship between eccentricity and surface temperature.
 
The HZ is usually conceived as a circular annulus surrounding the star. A planet can have a semi-major axis firmly within the limits, yet can make excursions beyond those limits because of non-zero eccentricity. It is not immediately obvious whether or not such excursions render the planet uninhabitable, because the atmosphere and ocean buffer the heat loss or gain. \cite{williams2002} used a 3D GCM and a 1D EBM to look at the effects of large eccentricity and concluded that seasonal excursions beyond the limits of the HZ are survivable even up to an eccentricity of $\sim 0.7$. More recent GCM work by \cite{bolmont2016} similarly found that excursions beyond the HZ are not typically fatal, however, they found that at $e\sim0.6$ in the HZs of hotter stars, planets spend so much time near apoastron that snowball events occur. 

Using the same EBM as \cite{spiegel2009}, \cite{dressing2010} showed the stronger dependence of the HZ boundaries on eccentricity, finding that the outer edge of the HZ can be pushed out by up $0.8$ au at $e = 0.5$, for example, for a planet orbiting a solar-type star. These authors did not see snowball event occurring at apoastron, even up to eccentricity of $0.9$.

The first study to look at how orbital \emph{variations}, or exo-Milankovitch cycles, affect climate for exoplanets was \cite{spiegel2010}. These authors focused on the eccentricity cycles induced by having a giant planet perturbing a rocky HZ planet. Notably, they found that a frozen planet could be deglaciated if the eccentricity is excited to large enough values. 

To further explore the dynamical influence of climate, \cite{armstrong2014} constructed various planetary systems which included an Earth-mass HZ planet, and one or two companions of various masses. To our knowledge, the first study to bring orbital dynamics, obliquity evolution, and climate modeling together for exoplanets, they modeled the evolution of these systems over 1 Myr, and then used the resulting orbital and obliquity output as initial conditions for a simple 1D EBM. By also including the growth and decay of ice sheets (\emph{i.e.} quasi-permanent ice on land), they found that the outer edge of the HZ depends on the orbital evolution. One shortcoming of the study was that the EBM used did not include latitudinal heat diffusion, which was shown by \cite{spiegel2008} and \cite{spiegel2009} to be very important. Thus their results are best viewed as a proof of concept, rather than an exact determination of the outer limit of the HZ. That study also did not include a detailed examination of how the large amplitude obliquity oscillations arise.
 
\cite{atobe2004} looked at obliquity evolution of hypothetical Earth-sized planets in systems with known giant planets. They used a second order orbital theory to map the habitable zones in terms of the severity of obliquity oscillations, assuming coplanarity. In particular, they noted the importance of secular spin-orbit resonances in triggering large amplitude variations. However, \cite{atobe2004} did not consider the effects of higher order orbital theory and of mutual inclinations.

One of the first studies to apply the concept of exo-Milankovitch cycles to known exoplanets, \cite{brasser2014} modeled the dynamics of the planetary system orbiting HD 40307 and calculated the resulting insolation that would be received by planet g. This outer planet, if it exists \citep[see][]{diaz2016}, is probably a gaseous mini-Neptune \citep{rogers2015}, rather than a rocky super-Earth, but nevertheless \cite{brasser2014} remains an important study of the potential effects that dynamics can have on habitability. They noted a secular resonance which, depending on planet g's initial obliquity and rotation rate, could strongly influence the obliquity evolution and thus the insolation received at each latitude. These authors did not apply a climate model, however. 

More recently, \cite{forgan2016} looked at Milankovitch cycles for several planets in binary systems with an EBM similar to that used by \cite{spiegel2009}. The presence of a second star can induce large changes in the surface temperatures, particularly for S-type planets (those that orbit a single star within a binary). This problem warrants further study, particularly because surface albedos depend very strongly on the spectral type of the host stars \citep{shields2013}, so that the stellar fluxes of each star may need to be treated differently in the climate model.
 
Finally, \cite{shields2016} modeled the dynamics, including tidal forces, of the Kepler-62 planets and the possible climate of planet f. Kepler-62 f lies in the outer part of the star's HZ, according to the calculations of \cite{kopparapu2013}, and so an increased abundance of CO$_2$ or some other greenhouse gas is necessary to keep the planet from freezing. In addition to demonstrating the stability of the system, these authors also modeled the climate of planet f using two different 3D GCMs and found that with its mean eccentricity, the planet can be habitable but that this depends strongly on its obliquity and atmospheric CO$_2$ pressure.

In this work, we study the orbital and obliquity evolution of Kepler-62 f, HD 40307 g (briefly), and a mutually-inclined Earth--Jovian system. Previous studies have looked at the orbital evolution and obliquity damping (due to tides) of Kepler-62 f \citep{bolmont2015,shields2016}, but have not explored the obliquity evolution due to gravitational perturbations. And while \cite{armstrong2014} explored the effect of mutual inclinations in a handful of configurations, an in-depth mapping of parameter space for a mutually inclined system, including different initial rotation states, has not yet been done. 

The goals of this study are (1) to introduce our secular (orbit-averaged) orbital framework, which allows us to map wide regions of parameter space; (2) to demonstrate the complexity of planetary systems and the potential ways orbital evolution can influence climate, including large eccentricities and mutual inclinations, in preparation for the discovery of dynamically ``hot'' systems; and (3) to generate initial conditions for climate modeling in a follow up study (Deitrick et al., in prep). Our formulations allow us to understand the fundamental causes of large obliquity variations, and to identify higher order secular resonances that depend on eccentricity and inclination.

Section \ref{sec:methods} contains a description of our model, which we validate against an N-body model in Section \ref{sec:valid}. In Section \ref{sec:results}, we lay out results from the model for several example systems (Kepler-62, HD 40307, and a test system), and discuss secular resonances and Cassini states. In Section \ref{sec:discuss}, we offer some interpretation and broader context, and we conclude in Section \ref{sec:conclu}. Finally, the Appendix contains the disturbing function used in the orbital model, written in terms of our integration variables. 

\section{METHODS}
\label{sec:methods}

\subsection{Orbital model}
\label{distorb}
Our model for the orbital evolution, called \texttt{DISTORB} (for ``Disturbing function Orbits''), uses the literal, 4th order disturbing function developed in \cite{md1999} and \cite{ellis2000}. We use only the secular terms, meaning that the rapidly varying terms that depend on the mean longitudes of the planets are ignored on the assumption that these terms will average to zero over long timescales. This is a valid assumption as long as no planets are in proximity of mean-motion resonances. There are, in fact, two orbital models in \texttt{DISTORB}: the first is simply a direct Runge-Kutta integration of the fourth-order equations of motion with variable time-stepping; the second is the Laplace-Lagrange eigenvalue solution, which reduces the accuracy in the disturbing function to second-order, but returns a solution that is explicit in time, and thus provides a solution in much less computation time. Most of the work we present here takes advantage of the fourth-order solution for its accuracy, but we discuss some results using the Laplace-Lagrange method in Sections \ref{sectsys2} and \ref{cassini}.

The equations of motion are Lagrange's equations \citep[see][]{md1999}. In the secular approximation the equations for semi-major axis and mean longitude, and any disturbing function derivative with respect to these variables, are ignored. Additionally, to avoid singularities in the equations for the longitudes of pericenter and ascending node which occur at zero eccentricity and inclination, respectively, we rewrite Lagrange's equations and the disturbing function in terms of the variables (a form of Poincar\'{e} coordinates):
\begin{align}
h & = e \sin{\varpi} \\
k & = e \cos{\varpi} \\
p & = \sin{\frac{i}{2}} \sin{\Omega} \label{eqnp}\\
q & = \sin{\frac{i}{2}} \cos{\Omega} \label{eqnq},
\end{align}
where $e$ is the orbital eccentricity, $i$ is the inclination, $\Omega$ is the longitude of ascending node, and $\varpi = \omega+\Omega$ is the longitude of periastron (see Figure \ref{diagpA}).
 
Lagrange's equations for secular theory are then:
\begin{align}
&\begin{aligned}
\frac{dh}{dt} & = \frac{\sqrt{1-e^2}}{na^2} \frac{\partial\mathcal{R}}{\partial k} + \frac{k p}{2na^2\sqrt{1-e^2}} \frac{\partial\mathcal{R}}{\partial p} \\
  &\qquad + \frac{k q}{2na^2\sqrt{1-e^2}} \frac{\partial\mathcal{R}}{\partial q}
\end{aligned}\label{eqnhdot}\\
&\begin{aligned}
\frac{dk}{dt} & = -\frac{\sqrt{1-e^2}}{na^2} \frac{\partial\mathcal{R}}{\partial h} - \frac{h p}{2na^2\sqrt{1-e^2}} \frac{\partial\mathcal{R}}{\partial p} \\
&\qquad-\frac{h q}{2na^2\sqrt{1-e^2}} \frac{\partial\mathcal{R}}{\partial q}
\end{aligned}\label{eqnkdot}\\
&\begin{aligned}
\frac{dp}{dt} & = - \frac{k p}{2na^2\sqrt{1-e^2}} \frac{\partial\mathcal{R}}{\partial h} +\frac{h p}{2na^2\sqrt{1-e^2}} \frac{\partial\mathcal{R}}{\partial k} \\
&\qquad + \frac{1}{4na^2\sqrt{1-e^2}} \frac{\partial\mathcal{R}}{\partial q} 
\end{aligned}\label{eqnpdot}\\
&\begin{aligned}
\frac{dq}{dt} & = - \frac{k q}{2na^2\sqrt{1-e^2}} \frac{\partial\mathcal{R}}{\partial h} +\frac{h q}{2na^2\sqrt{1-e^2}} \frac{\partial\mathcal{R}}{\partial k}\\
&\qquad - \frac{1}{4na^2\sqrt{1-e^2}} \frac{\partial\mathcal{R}}{\partial p},
\end{aligned} \label{eqnqdot}
\end{align}
where $\mathcal{R}$ is the disturbing function (see Appendix), and $a$, $n$, and $e$ are the semi-major axis, mean motion, and eccentricity, respectively. See \cite{berger1991} for the complete set of Lagrange's equations in $h$, $k$, $p$, and $q$, including mean-motion (\emph{e.g.}, resonant) effects. 

General relativity is known to affect the apsidal precession (associated with eccentricity) of planetary orbits, so we include a correction to Equations (\ref{eqnhdot}) and (\ref{eqnkdot}). Following \cite{laskar1986}, the apsidal corrections are:
\begin{align}
\frac{dh}{dt}\Bigm\lvert_{GR} &= \delta_R k \label{hGRcorr} \\ 
\frac{dk}{dt}\Bigm\lvert_{GR}& = -\delta_R h,\label{kGRcorr}
\end{align}
where 
\begin{equation}
\delta_R = \frac{3 n^3 a^2}{c^2 (1-e^2)},
\end{equation}
and $c$ is the speed of light. 

The secular approximation allows us to take large time-steps (years to hundreds of years) and thus to run thousands of simulations quickly and explore parameter space with relatively minimal computer usage compared with N-body models.

\subsection{Obliquity model}
\begin{figure}[t]
\includegraphics[width=0.5\textwidth]{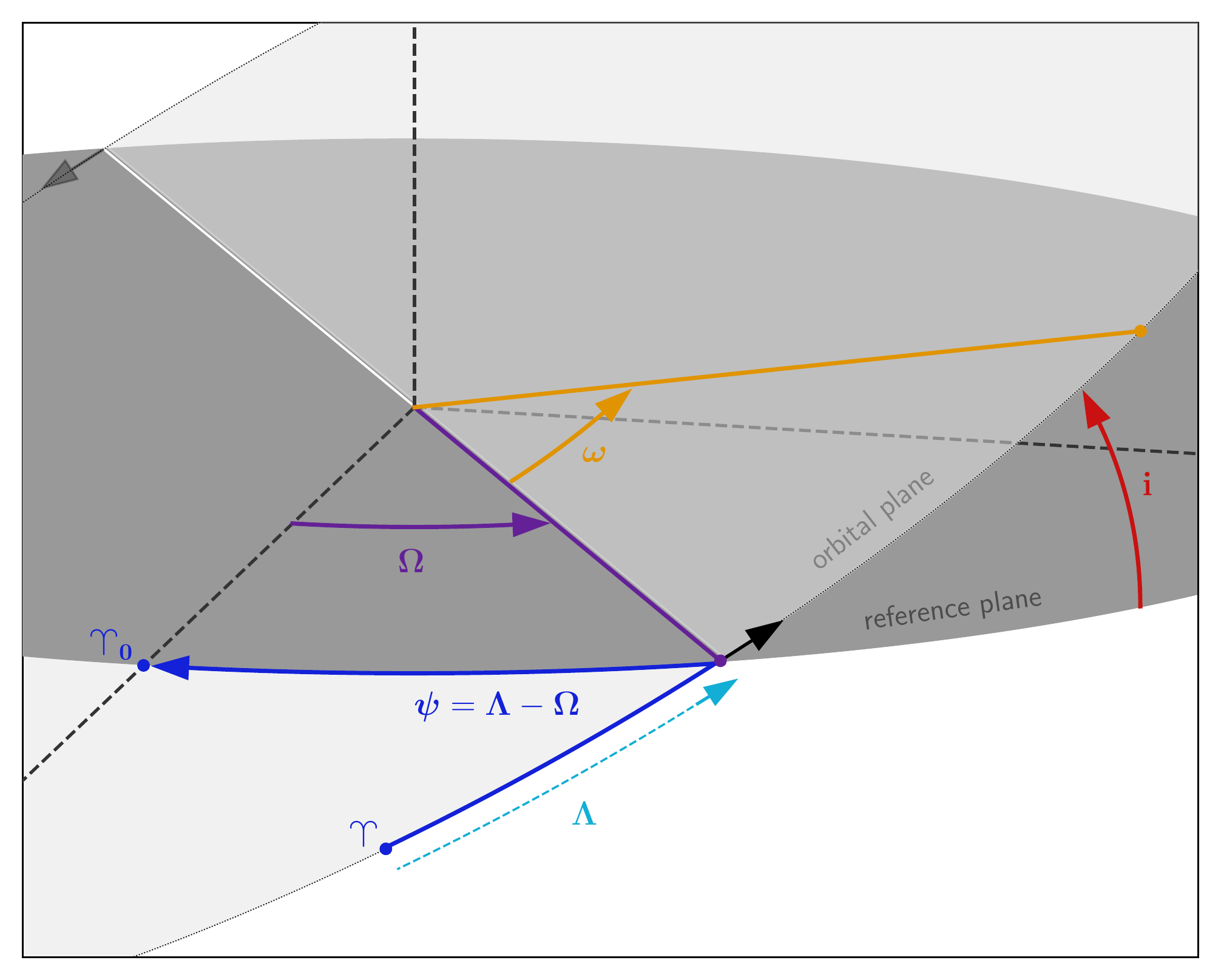}
\caption{\label{diagpA} Geometry used in the obliquity model, \texttt{DISTROT}. The light gray represents the planet's orbital plane, while the darker gray represents a plane of reference. The important orbital angles are the inclination, $i$, the longitude of ascending node, $\Omega$, and the argument of pericenter, $\omega$. The \emph{longitude} of pericenter is a ``dog-leg'' angle, $\varpi = \Omega + \omega$. The angle $\Lambda$ is measured from the vernal point $\vernal$ at time $t$, to the ascending node, $\Omega$. The precession angle is defined as $\psi = \Lambda - \Omega$ (also a dog-leg angle). The reference point for $\Omega$ is usually chosen as the vernal point at some known date for solar system, however, there is probably a more sensible choice for exoplanetary systems.}
\end{figure}

The obliquity model, \texttt{DISTROT} (for ``Disturbing function Rotation''), is derived from the Hamiltonian for rigid body motion introduced by \cite{kino1975, kino1977} and later used by \cite{laskar1986, laskar1993, armstrong2004, armstrong2014} and several others. In the absence of large satellites (such as the Moon), the equations of motion for a rigid planet are:
\begin{align}
&\begin{aligned}
\frac{d\psi}{dt} & = R(\varepsilon) - \cot(\varepsilon)~[A(p,q) \sin{\psi} \\ 
 &\qquad +  B(p,q) \cos{\psi}] - 2\Gamma(p,q) - p_g
\end{aligned}\label{eqnpA}\\
&\begin{aligned}
\frac{d\varepsilon}{dt} & = -B(p,q) \sin{\psi} + A(p,q) \cos{\psi} \label{eqnpsi},
\end{aligned}
\end{align}
where $\psi $ is the ``precession angle'' (see Figure \ref{diagpA} and the following paragraph), $\varepsilon$ is the obliquity, and,
\begin{align}
R(\varepsilon) & = \frac{3 \kappa^2 M_{\star}}{a^3 \nu} \frac{J_2 M r^2}{C} S_0 \cos{\varepsilon} \label{eqnR}\\
S_0 & = \frac{1}{2} (1-e^2)^{-3/2}  \\
A(p,q) & = \frac{2}{\sqrt{1-p^2-q^2}} [\dot{q}+p \Gamma(p,q)] \label{eqnA}\\
B(p,q) & = \frac{2}{\sqrt{1-p^2-q^2}} [\dot{p}-q \Gamma(p,q)] \label{eqnB} \\
\Gamma(p,q) & = q\dot{p}-p\dot{q}. \label{eqnC}
\end{align}
Note the sign error in Equation (8) of \cite{armstrong2014}, corrected in our Equation (\ref{eqnA}). Our Equation (\ref{eqnR}) does not contain the lunar constants present in Equation (24) of \cite{laskar1986}. Here, $\varepsilon$ represents the obliquity, $p$ and $q$ as in Equations (\ref{eqnp}) and (\ref{eqnq}), $\dot{p}$ and $\dot{q}$ are their time derivatives (Equations (\ref{eqnpdot}) and (\ref{eqnqdot})), $\kappa$ is the Gaussian gravitational constant, $M_{\star}$ is the mass of the host star in solar units, $\nu$ is the rotation frequency of the planet in rad day$^{-1}$, $C M^{-1}r^{-2}$ is the specific polar moment of inertia of the planet, and $J_2$ is the gravitational quadrupole of the (oblate) planet.

The final term in Equation (\ref{eqnpA}), $p_g$, accounts for precession due to general relativity and is equal to $\delta_R/2$ \citep{barker1970}. The symbol $\psi$ refers to the precession angle, defined as $\psi = \Lambda - \Omega$, where $\Lambda$ is the angle between the vernal point $\vernal$, the position of the sun/host star at the planet's northern spring equinox, and the location of the ascending node, $\Omega$, measured from some reference direction $\vernal_0$ (often taken to be the direction of the vernal point at some reference date, hence the use of the symbol $\vernal$\footnote[1]{The vernal point occurred in the constellation Aries during Ptolemy's time, thus it is also called the ``first point of Aries'' and the ``ram's horn'' symbol is used.}), see Figure \ref{diagpA}. The convention of defining $\vernal$ as the location of the sun at northern spring equinox is sensible for the solar system since we observe from Earth's surface, however, it is a confusing definition to use for exoplanets, for which the direction of the rotation axis is unknown. We adhere to the convention for the sake of consistency with prior literature. In the coming decades, it may become possible to determine the obliquity and orientation of an exoplanet's spin axis; in that event, care should be taken in determining the initial $\psi$ for obliquity modeling. One can equivalently refer to $\vernal$ as the position of the planet at its northern spring equinox $\pm 180^{\circ}$. The relevant quantity for determining the insolation, however, is the angle between periastron and the spring equinox, $\omega + \Lambda = \varpi + \psi$.

An additional complication for obliquity evolution is, of course, the presence of a large moon. We do not include the component of the \cite{kino1975} model that accounts for the lunar torque because the coefficients used are specific to the Earth-Moon-Sun three body problem and were calculated from the Moon's orbital evolution (and are therefore not easily generalized). However, we can approximate the effect of the Moon by \emph{forcing} the precession rate (Equation (\ref{eqnR})) to be equal to the observed terrestrial value. We do not use this feature of the model in our exploration of exoplanets, but use it to validate our models by reproducing the Earth's obliquity evolution. 

Equation (\ref{eqnpA}) for the precession angle contains a singularity at $\varepsilon = 0$. To avoid numerical instability, we instead recast Equations (\ref{eqnpA}) and (\ref{eqnpsi}) in terms of the rectangular coordinates:
\begin{align}
\xi & = \sin{\varepsilon} \sin{\psi} \\
\zeta & = \sin{\varepsilon} \cos{\psi} \\
\chi & = \cos{\varepsilon}.
\end{align}
The third coordinate, $\chi$, is necessary to preserve sign information when the obliquity crosses 90$^{\circ}$ \citep[see][]{laskar1993}. The equations of motion for these variables are then:
\begin{align}
\frac{d\xi}{dt} & = -B(p,q)\sqrt{1-\xi^2-\zeta^2} + \zeta[R(\varepsilon)-2\Gamma(p,q)-p_g] \\
\frac{d\zeta}{dt} & = A(p,q)\sqrt{1-\xi^2-\zeta^2} - \xi[R(\varepsilon)-2\Gamma(p,q)-p_g] \\
\frac{d\chi}{dt} & = \xi B(p,q) - \zeta A(p,q).
\end{align}

The value of $J_2$ is a function of the planet's rotation rate and density structure. In hydrostatic equilibrium, the Darwin-Radau relation \citep{cook1980, md1999} shows that the planet's gravitational quadrupole moment, $J_2$, scales with the rotation rate squared. Earth is very close to hydrostatic equilibrium, while Mars and the Moon are not. For the purposes of this study, we assume that planets are in hydrostatic equilibrium, and scale $J_2$ according to:
\begin{equation}
J_2 = J_{2\oplus} \left (\frac{\nu}{\nu_{\oplus}} \right)^2 \left (\frac{R}{R_{\oplus}} \right)^3 \left (\frac{M}{M_{\oplus}} \right)^{-1} \label{eqnJ2},
\end{equation}
where $\nu$ is the rotation rate, $R$ is the mean planetary radius, and $M$ is its mass, and Earth's measured values of $J_{2\oplus} = 1.08265 \times 10^{-3}$, $\nu_{\oplus}=7.292115 \times 10^{-5}$ radians s$^{-1}$, $R_{\oplus} =  6.3781\times 10^{6}$ m, and $M_{\oplus}= 5.972186\times 10^{24}$ kg are used for reference ($J_2$ from \cite{cook1980}, $R_{\oplus}$ and $M_{\oplus}$ from \cite{prsa2016}). This is identical to the method used by \cite{brasser2014}. Like that study, we assume that $J_2$ cannot go below the measured value of Venus from \cite{yoder1995}, which would ordinarily occur at rotation periods $\gtrsim 13$ days. The assumption then is that there is a limit to hydrostatic equilibrium for a partially rigid body, which is difficult to validate, but does not affect our results greatly as most of our parameter space results in $J_2$ well above this value. 

The \emph{specific} polar moment of inertia of a planet, $C M^{-1} r^{-2}$ is always between 0.2 and 0.4.
Here we assume the $C M^{-1} r^{-2}$ value of the Earth, $0.33$, \citep{cook1980}. A different assumption for $C M^{-1} r^{-2}$ \emph{will} change the exact values of our results, however, as the moment of inertia should be similar to this value for a terrestrial planet, the overall nature of our results shouldn't change dramatically. 

In Section \ref{sec:results}, we employ power spectra to identify resonances between the inclination and obliquity variables. The variables used are 
$\zeta + \sqrt{-1}\xi$ for the obliquity variations and $q+\sqrt{-1}p$ for the inclination variations. We use the \texttt{Python} function \texttt{periodogram} from the package \texttt{scipy.signal} \citep{jones2001} to calculate these power spectra. This function performs fast-Fourier transforms that reduce signal leakage through Welch's algorithm \citep{welch1967} and use a window function to produce a cleaner spectrum. For our study, we use a Hanning window function.

\subsection{Cassini state theory}
\label{casstheory}
The coupling of the obliquity model and the orbital model allows for the appearance of Cassini states, and indeed, we do see several of their features in our simulations. These special rotation states will have implications for obliquity evolution and climate, and so it is worth reviewing the theory here. 

A planet or satellite is in a Cassini state if Cassini's laws (initially described in reference to the Moon) are satisfied. Cassini's laws, generalized and paraphrased here, are:
1. The planet or satellite is in a spin-orbit resonance, such as the synchronous rotation of the Moon, or the 3:2 resonance of Mercury.
2. The planet's or satellite's obliquity is approximately constant, with respect to an appropriate plane of reference, such as the invariable plane (the plane perpendicular to the total angular momentum) of the planetary system or the Laplace plane of the satellite (defined by the host planet's orbit and equator). 
3. Three vectors, the spin angular momentum of the planet or satellite, its orbital angular momentum, and the perpendicular to the reference plane from the second law, are approximately coplanar at all times. 

Though Cassini wrote these laws in reference to Earth's Moon, these were later generalized by \cite{colombo1966} and \cite{peale1969} and found to apply to Mercury and other natural satellites of the solar system. The Cassini state theory developed by these later authors is elegant and mathematically rigorous and can be used to describe the trajectory of a body's spin axis as it evolves toward or around any of four unique Cassini states. The theory can be applied when the body is not actually within one of the Cassini states, provided that the obliquity evolution is dominated by the central torque and a single perturbing frequency \citep[see][]{wardhamilton2004,hamiltonward2004}. When the theory applies, the body's spin axis will oscillate about one of the Cassini states and Cassini's third law may be satisfied, even when the first two are not. A damping force, such as tidal friction, can drive the body into its ultimate state which satisfies all three of Cassini's laws, hence the prevalence of Cassini states among the solar system satellites. 

The conditions specified by the first two laws are easily identified in dynamical simulations of obliquity and orbital evolution. The third law can be identified by invoking the following relation, from \cite{hamiltonward2004}:
\begin{equation}
\sin{\Psi} = \left \| \frac{(\mathbf{k} \times \mathbf{n})\times(\mathbf{s}\times\mathbf{n})}{|\mathbf{k} \times \mathbf{n}||\mathbf{s}\times\mathbf{n}|} \right \|, \label{cass1}
\end{equation}
where $\mathbf{k}$, $\mathbf{n}$, and $\mathbf{s}$ are the vectors associated with the perpendicular to the appropriate reference plane (the invariable plane or Laplace plane, for example), the angular momentum of the body's orbit, and the angular momentum of the body's rotation. Alternatively, the complimentary relation can be used:
\begin{equation}
\cos{\Psi} = \frac{(\mathbf{k} \times \mathbf{n})\cdot(\mathbf{s}\times\mathbf{n})}{|\mathbf{k} \times \mathbf{n}||\mathbf{s}\times\mathbf{n}|}. \label{cass2}
\end{equation}

Close inspection indicates that $\Psi = \Lambda = \psi + \Omega$ (when $\Omega$ is measured in the invariable plane of the system), and Cassini's third law is satisfied when this angle librates about $0^{\circ}$ or $180^{\circ}$ instead of circulating. Since a secular spin-orbit resonance occurs when the axial precession frequency is similar to and opposite in sign to a nodal precession frequency, libration of this angle is also indicative of a strong resonance. A Cassini state is thus a form of secular spin-orbit resonance, though this type of resonance can still occur when not all of Cassini's laws are satisfied.

The locations of Cassini states can be determined from the relation \citep{wardhamilton2004}:
\begin{equation}
(\alpha/s_i) \cos{\theta_j} \sin{\theta_j} + \sin{(\theta_j-I_{i})} = 0,
\end{equation}
where $\theta_j$ is the angle between the spin-axis vector and a normal to the planet's orbital plane for the $j$th Cassini state\footnote{If $\theta_j$ measured from $-180^{\circ} < \theta_j<180^{\circ}$, $| \theta_j |$ is equal to the obliquity, $\varepsilon$}, $s_i$ is the $i$th eigenvalue associated with the Laplace-Lagrange solution for inclination, $I_i$ is the corresponding amplitude of the oscillation in inclination, and $\alpha$ is the precessional constant, 
\begin{equation}
\alpha = \frac{R(\varepsilon)}{\cos{\varepsilon}} = \frac{3 \kappa^2 M_{\star}}{a^3 \nu} \frac{J_2 M r^2}{C} S_0.  \label{eqn:alpha}
\end{equation}

This equation can be solved exactly using complex algebra to rewrite it as a quartic polynomial, but the approximate solutions given by \cite{wardhamilton2004} are actually very accurate. They are:
\begin{equation}
\theta_{1,3} \approx \tan^{-1} \left( \frac{\sin{I_i}}{1\pm \alpha/s_i} \right),
\label{eqn:cass13}
\end{equation}
and
\begin{equation}
\theta_{2,4} \approx \pm \cos^{-1} \left( \frac{-s_i \cos{I_i}}{\alpha} \right).
\label{eqn:cass24}
\end{equation}

Cassini state 1 tends to occur at $| \theta_i | = \varepsilon \approx 0^{\circ}$ and state 3 at $| \theta_i | = \varepsilon \approx 180^{\circ}$ (retrograde spin). States 2 and 4 have the same $| \theta_i | = \varepsilon$, but the precession angles differ by $180^{\circ}$ ($\psi_4 = \psi_2 \pm 180^{\circ}$), \emph{i.e.}, the spin vectors are on opposite sides of the orbit normal. State 4 is a saddle point, as shown in Figure 2 in \cite{wardhamilton2004}, and is therefore unstable. Returning to the angle $\Psi = \psi + \Omega$: in the vicinity of Cassini state 2, $\psi +\Omega$ librates about zero; in the vicinity of Cassini state 4, $\psi + \Omega$ circulates or librates with large amplitude (\emph{i.e.}, $\sim 360^{\circ}$) in a ``horseshoe'' trajectory.

\subsection{Initial conditions}

The first planetary system we explore is that of Kepler-62, a K type star with 5 known transiting planets \citep{borucki2011,borucki2013}. The outer-most planet, f, is particularly interesting for the purposes of this study because it is in the habitable zone and far enough from the star to potentially have any spin-state (as opposed to tidally-locked and zero obliquity) \citep{bolmont2015}. We use two sets of masses (the stable case from \cite{bolmont2015} and the case from \cite{kopparapu2014}) and eccentricities and inclinations from \cite{borucki2013}, varying the obliquity, rotation rate, and precession angle of planet f. The longitude of pericenter, $\varpi$, was randomly chosen for each planet, while the longitude of ascending node, $\Omega$, was set to zero to keep the planets very coplanar. The orbital and spin properties are given in Table \ref{k62table}---the inclination, longitude of ascending node, and longitude of periastron are listed with respect to the invariable (angular momentum) plane of the system  ($i_{inv}$, $\Omega_{inv}$, and $\varpi_{inv}$) and the sky plane ($i_{sky}$, $\Omega_{sky},$ and $\varpi_{sky}$). Since \cite{bolmont2015} showed that general relativistic corrections are important to the stability of this system, we include those here according to Equations (\ref{hGRcorr}) and (\ref{kGRcorr}).

\begin{table}[h]
\centering
\caption{Initial conditions for Kepler-62}

\begin{tabular}{lcccccc}
\hline\hline \\ [-1.5ex]
Planet & & b & c & d & e & f\\ [0.5ex]
\hline \\ [-1.5ex]

$m$ ($M_{\oplus}$)  & $\mathcal{A}$ & 2.72 & 0.136 & 14 &6.324 & 3.648\\
& $\mathcal{B}$ & 2.3 & 0.1 & 8.2 & 4.4 & 2.9\\
$a$ (au) & & 0.0553 & 0.0929 & 0.12 & 0.427 & 0.718 \\
$e$ &   & 0.071 & 0.187 & 0.095 & 0.13 & 0.094 \\
$i_{inv}$ ($^{\circ}$) & & 0.6138 & 0.1138 & 0.1138 & 0.1662 & 0.08615 \\
$\varpi_{inv}$ ($^{\circ}$) & & 268.175 & 228.074 & 241.127 & 42.615 & 6.277 \\
$\Omega_{inv}$ ($^{\circ}$) & & 270.002 &  270.011 & 270.012 & 89.992 & 89.984\\
$P_{rot}$ (days) & & & & & & 0.25-20  \\
$\varepsilon$ ($^{\circ}$) & & & & & & 0-90  \\
$\psi+ \Omega$ ($^{\circ}$) & & & &  & &0,90,180,270 \\
\hline
$i_{sky}$ ($^{\circ}$) & & 89.2 & 89.7 & 89.7 & 89.98 & 89.9 \\
$\varpi_{sky}$ ($^{\circ}$) & & 178.175 & 138.074 &151.127 & 312.615 & 276.2767\\
$\Omega_{sky}$ ($^{\circ}$) & & 0 & 0 & 0 & 0 & 0\\
\hline
\end{tabular}
\label{k62table}
\end{table}

The second system, HD 40307, we explore briefly because of its similarity (in some ways) to Kepler-62. The initial conditions are taken directly from Table 6 in \cite{brasser2014}. We vary the initial obliquity and rotation period of the outer planet, g, and examine its obliquity evolution. We run two sets of simulations: one with the initial value of $\psi+\Omega = 0^{\circ}$ (in the invariable plane of the system) and one with $\psi+\Omega = 180^{\circ}$.

Our final system, which we will refer to as TSYS, is one with a warm Neptune mass planet, an Earth-mass planet in the habitable zone, and a $1.5\text{ M}_{Jup}$ planet exterior to the HZ. This system is loosely based on HD 190360, which contains a Neptune-mass planet and the super-Jovian, and appears to have a stable habitable zone (assuming the minimum masses and that there are no additional planets between the two known planets). Though this is a fictitious system, we include it here in anticipation of discoveries that might be made by the \emph{PLATO} \citep{rauer2014} and \emph{Gaia} \citep{casertano2008, perryman2014} missions, and other future observatories. It is an extension of the work done by \cite{atobe2004, spiegel2010} and \cite{armstrong2014}, in that it is a test of an Earth-mass planet orbiting a sun-like star with a giant companion, which should not be ruled out as a possibility because it is undetectable by current observatories. Further, this system displays some interesting features that appear as a result of large mutual inclinations. The initial conditions and parameter space we explore for TSYS are shown in Table \ref{sys1table}. We fix the parameters of the two known planets and vary the initial eccentricity, inclination, rotation rate, and obliquity of the Earth-mass planet. The angle $\psi$ we explore briefly in $90^{\circ}$ intervals. For the sake of simplicity, we assume the planets 1 and 3 are roughly coplanar. 

\begin{table}[h]
\centering
\caption{Initial conditions for TSYS}

\begin{tabular}{lccc}
\hline\hline \\ [-1.5ex]
Planet & 1 & 2 & 3 \\ [0.5ex]
\hline \\ [-1.5ex]

$m$ ($M_{\oplus}$) & 18.75 & 1 & 487.81 \\
$a$ (au) & 0.1292 & 1.0031 & 3.973 \\
$e$ & 0.237 & 0.001-0.4 & 0.313 \\
$i$ ($^{\circ}$) & 1.9894 & 0.001-35 & 0.02126 \\
$\varpi$ ($^{\circ}$) & 353.23 & 100.22 & 181.13 \\
$\Omega$ ($^{\circ}$) & 347.70 & 88.22 & 227.95 \\
$P_{rot}$ (days) & & 0.1667-10 & \\
$\varepsilon$ ($^{\circ}$) & & 0-90 & \\
$\psi$ & & 281.78 & \\

\hline
\end{tabular}
\label{sys1table}
\end{table}

\section{MODEL VALIDATION}
\label{sec:valid}
\begin{figure*}[t]
\includegraphics[width=0.5\textwidth]{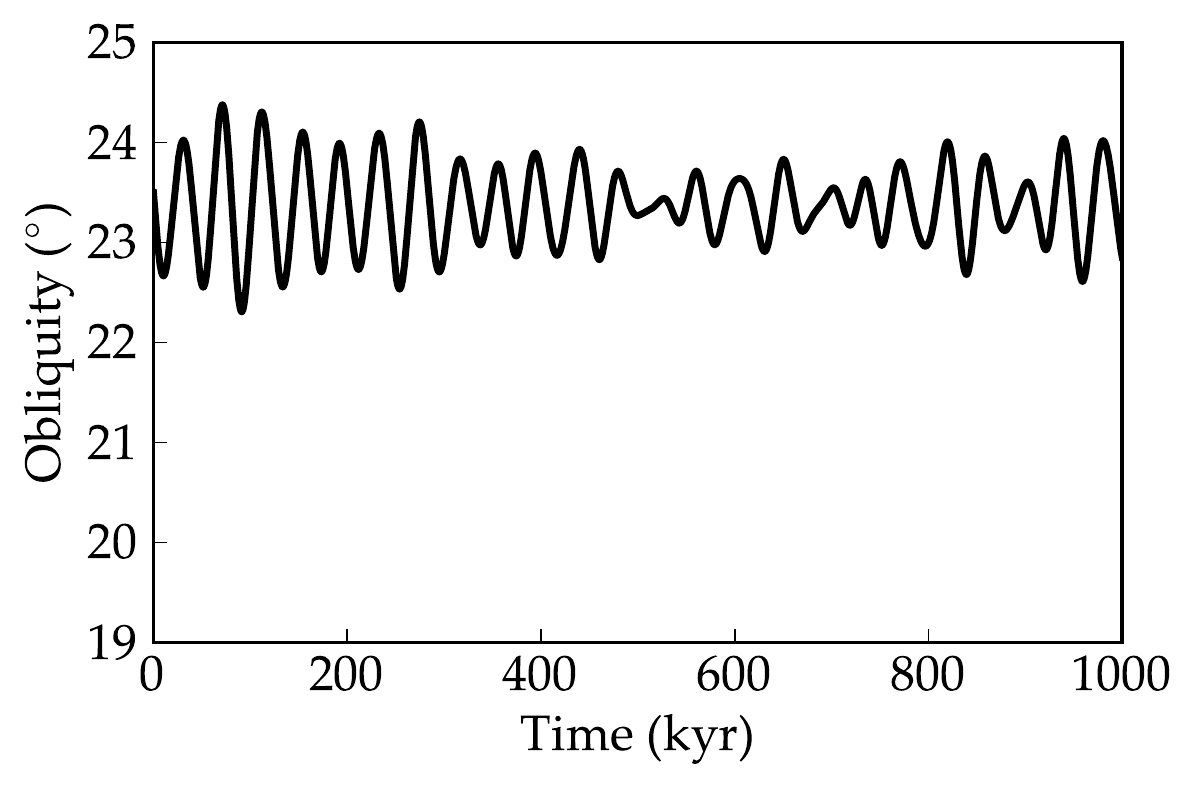}
\includegraphics[width=0.5\textwidth]{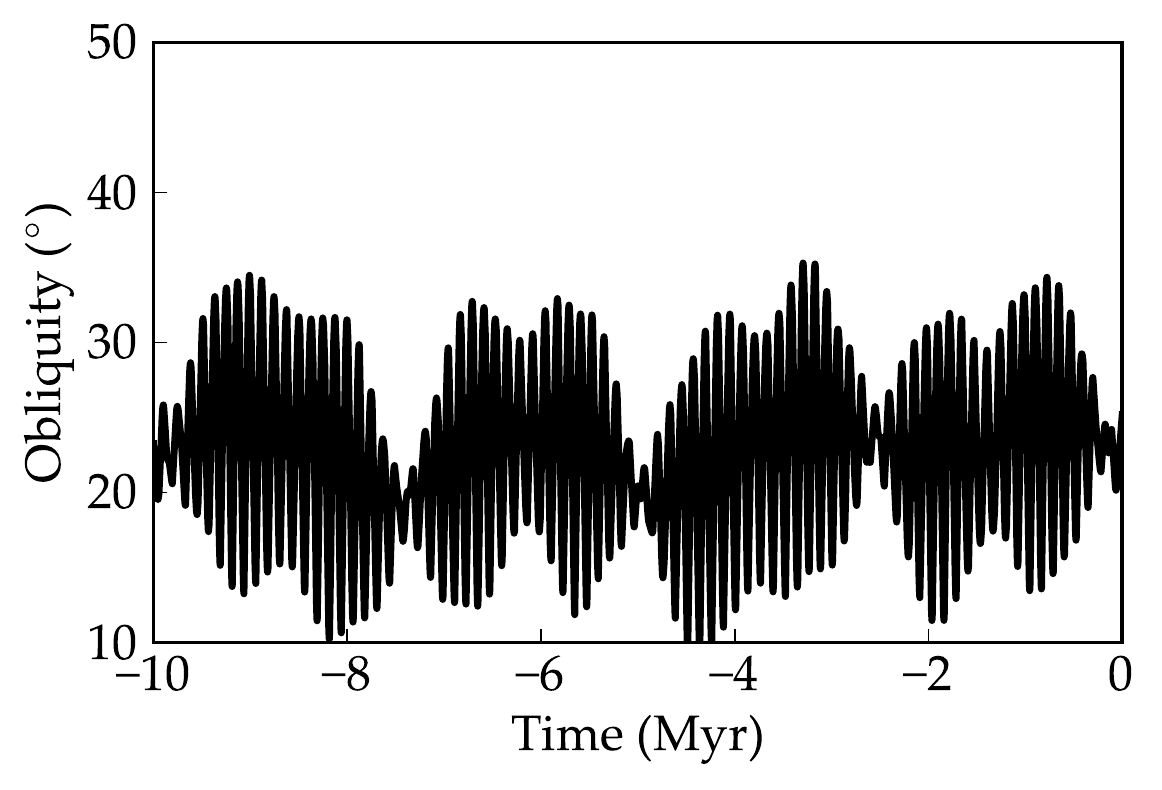}\\
\includegraphics[width=0.5\textwidth]{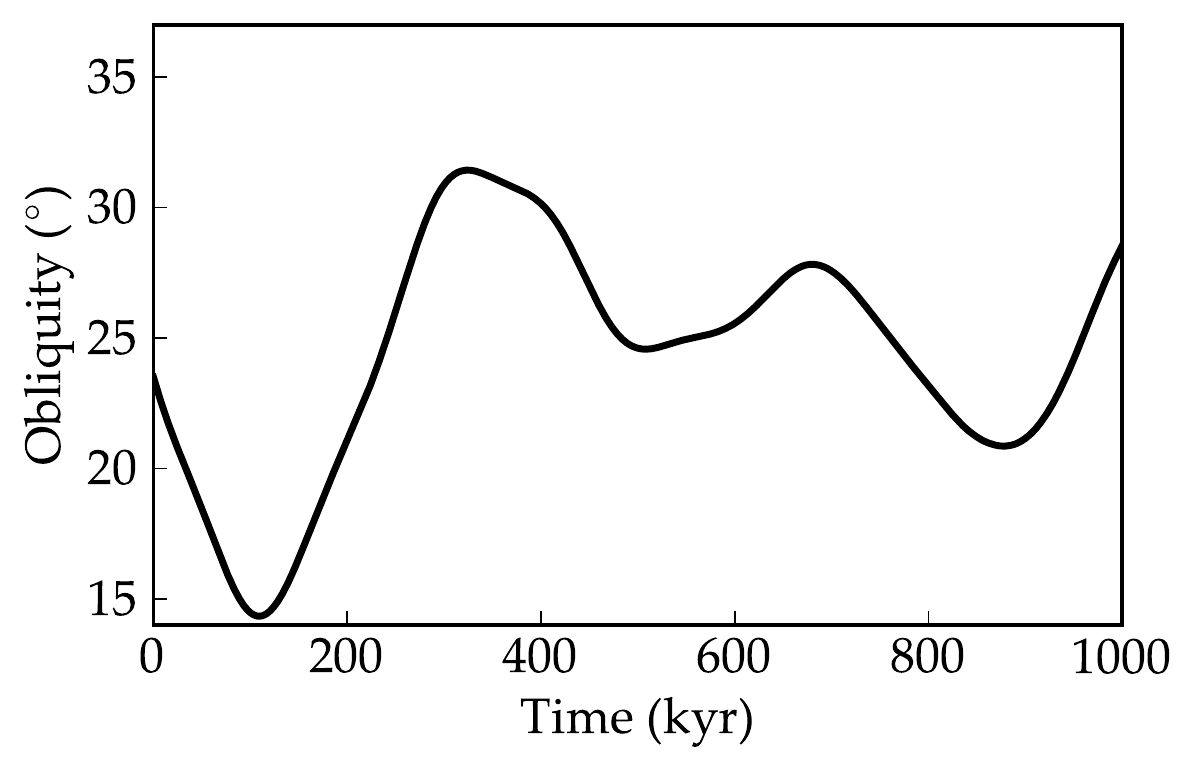}
\includegraphics[width=0.5\textwidth]{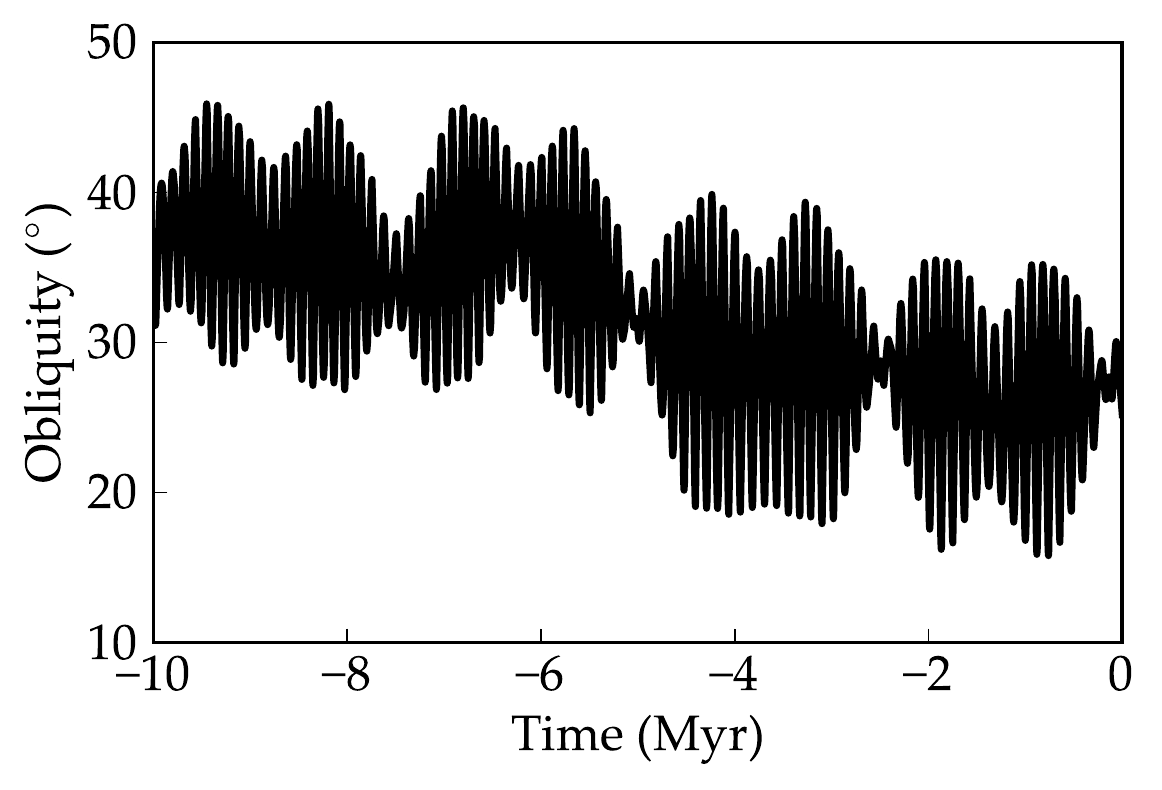}\\
\caption{\label{solarsysobl} Obliquity evolution for Earth over the next million years (upper left), from our secular model. Compare to Figure 10, top panel, in \cite{laskar1993}. Also shown is the obliquity evolution for Earth without the Moon (lower left), from our secular model. Compare to Figure 11, top panel, in \cite{laskar1993}. On the right, we have the obliquity evolution for Mars backwards in time with the secular orbital solution (upper right) and coupling \texttt{DISTROT} and \texttt{HNBody} (lower right).}
\end{figure*}

The orbital and obliquity models, \texttt{DISTORB} and \texttt{DISTROT} are validated by comparison with N-Body integrators and known results for the obliquity evolution of solar system bodies \citep{laskar1993,laskar1993b}. Figure \ref{solarsysobl} shows the resulting obliquity evolution for Earth with the Moon (upper left) and without (lower left). We simulate the effect of the Moon by forcing Earth's precession rate to equal the observed value of 50.290966 arcsec year$^{-1}$ \citep{laskar1993}. Both cases compare well with \cite{laskar1993}. This figure also shows the obliquity evolution for Mars in the past 10 Myr. The fully secular model captures the primary modes of oscillation and compares well with \cite{ward1992}. In the secular model, we do not see the transition to a higher obliquity regime at $\sim 5$ Myr ago seen in the simulations of \cite{touma1993} and \cite{laskar2004}. Mars' obliquity evolution is irregular and, as explained in \cite{touma1993}, is very sensitive to the details of the orbital model. The reason we do not reproduce this regime transition is most likely then because of the approximate nature of our orbital model. For studies of exoplanets, wherein the uncertainties in orbital parameters are much larger than for the solar system, our orbital model is still useful for broad parameter searches. We note that we can reproduce the transition at 5 Myr ago by coupling \texttt{HNBody} \cite{rauch2002} to our obliquity model (lower right panel).

\begin{figure*}[t]
\includegraphics[width=\textwidth]{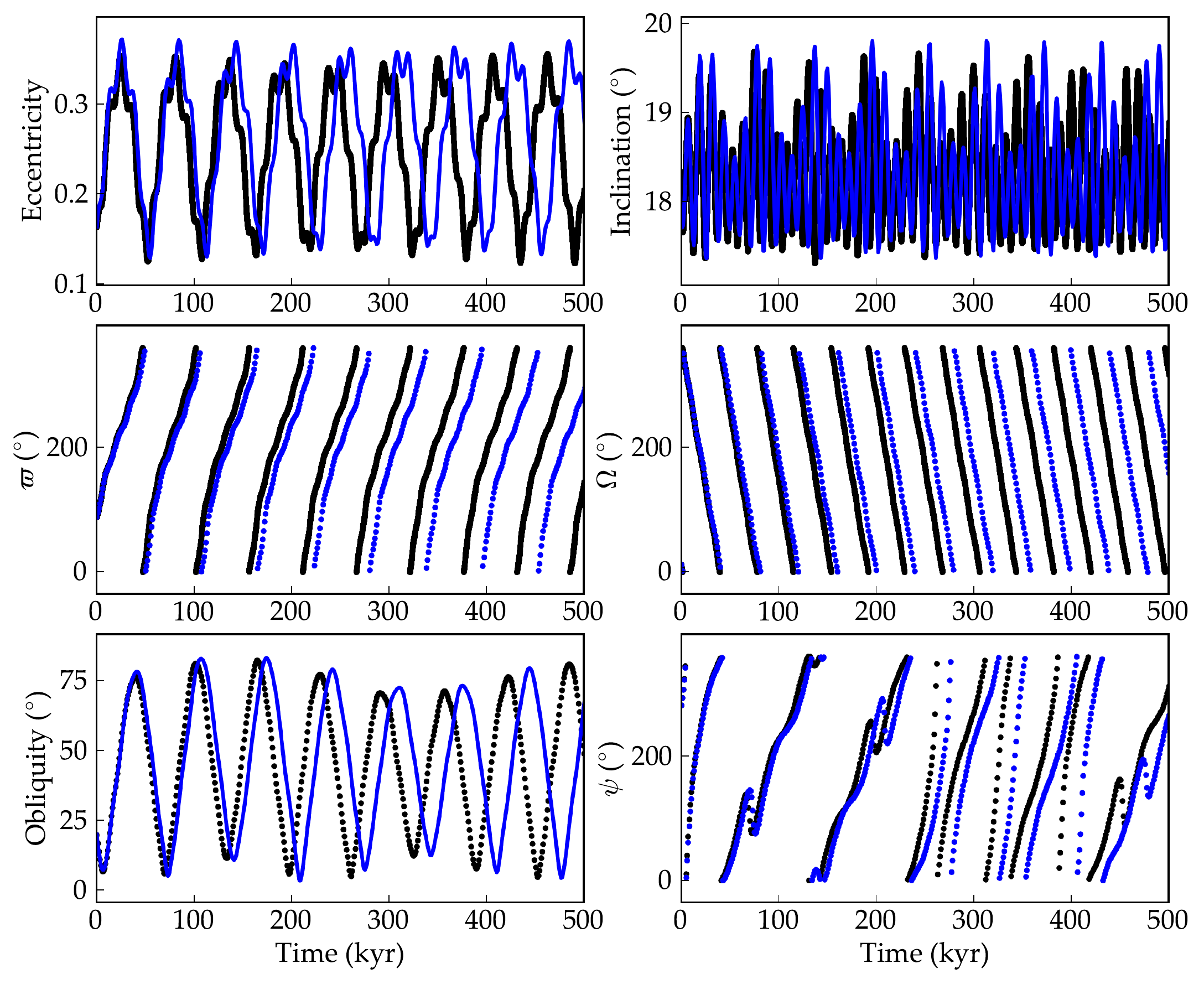}
\caption{\label{sys1nbody} Orbital and obliquity evolution for a case in our test system, TSYS (Section \ref{sectsys1}), comparing our secular model (blue) to an N-body model (black). There is some drift between the two solutions. Nevertheless, the secular model does an adequate job reproducing the general behavior of the system, in a small fraction of the computation time.}
\end{figure*}

One comparison with N-body models, using our system TSYS from Section \ref{sec:results}, is shown in Figure \ref{sys1nbody}. The N-body integration was done using \texttt{Mercury} \citep{chambers1999}. For the N-body results, the obliquity evolution was done using the code used in \cite{armstrong2014}, to provide a comparison to our new code \texttt{DISTROT}. 

The largest source of error between the secular models, \texttt{DISTORB} and \texttt{DISTROT}, comes from a slight offset in the frequencies, which we see as a drift in phase in Figure \ref{sys1nbody}. This is simply because some mean motion effects, such as minor variations in the semi-major axes of the planets, are neglected in the secular model. Overall, the secular model captures the general behavior (amplitudes of the oscillations and approximate frequencies) very well. The advantage of the secular model is that it can run millions of years in seconds, compared to hours with the N-body model, allowing us to explore wide regions of parameter space. When necessary or advantageous, we also perform the integrations using the N-body formulation.

\section{RESULTS}
We apply our model to a handful of simple planetary test systems, orbiting K and G stars, which reveal a wealth of dynamical (and potentially climatic) phenomena.
\label{sec:results}

\subsection{Kepler-62 f}
\label{sectsys2}

For our first system, Kepler-62, we vary the spin parameters of planet f, with the input parameters in Table \ref{k62table}. Because the eigenvalues from the Laplace-Lagrange solution are useful in understanding secular resonances and Cassini states, we calculate their values for both mass sets and both integration schemes (secular and N-Body). These are presented in Tables \ref{k62eign1} and \ref{k62eign2}. The eigenvalues $g_i$ are associated with the eccentricity evolution and $s_i$ with the inclination evolution.

\begin{table}[h]
\centering
\caption{Eigenvalues for Kepler-62 (secular solution)}

\begin{tabular}{l|cc|cc}
\hline\hline \\ [-1.5ex]
 & Mass set $\mathcal{A}$ & & Mass set $\mathcal{B}$ &\\
  & $g_i$ (arcsec year$^{-1}$) & $s_i$ (arcsec year$^{-1}$) & $g_i$ (arcsec year$^{-1}$) & $s_i$ (arcsec year$^{-1}$)\\
  \hline
$1$ & 4923.54 & -4932.41 & 2960.31 &-2966.71 \\
$2$& 659.46 & -681.49& 409.23 &-423.26\\
$3$  &76.54& -1.56$\times$10$^{-5}$& 61.05 &2.35$\times10^{-6}$\\
$4$ &53.24& -21.71& 38.16& -14.07\\
$5$  &16.92 &-63.27&11.02 &-45.17\\

\hline
\end{tabular}
\label{k62eign1}
\end{table}

\begin{table}[h]
\centering
\caption{Eigenvalues for Kepler-62 (N-body solution)}

\begin{tabular}{l|cc|cc}
\hline\hline \\ [-1.5ex]
& Mass set $\mathcal{A}$ & & Mass set $\mathcal{B}$ &\\
  & $g_i$ (arcsec year$^{-1}$)) & $s_i$ (arcsec year$^{-1}$))& $g_i$ (arcsec year$^{-1}$)) & $s_i$ (arcsec year$^{-1}$))\\
  \hline
$1$ & 4877.92 & -4886.79  & 2950.05 &-2956.55\\
$2$& 630.38 & -681.71  & 380.74 &-423.28\\
$3$  &75.07 & -1.74$\times$10$^{-15}$ & 58.94 & -1.90$\times10^{-16}$\\
$4$ &53.21 & -21.70 & 37.94 &-14.06\\
$5$  &16.90 &-63.28 & 11.32 &-45.10\\

\hline
\end{tabular}
\label{k62eign2}
\end{table}

In Table \ref{k62amplitude1}, we have listed the amplitudes of the inclination oscillations associated with each eigenvalue for planet f. These are calculated from the scaled eigenvectors of the Laplace-Lagrange solution. From this, it is clear that planet f's inclination is primarily driven by the frequency $s_4$, though $s_5$ also contributes strongly. As described below, these two frequencies will be responsible for secular resonances and Cassini states.

\begin{figure*}[t]
\includegraphics[width=0.5\textwidth]{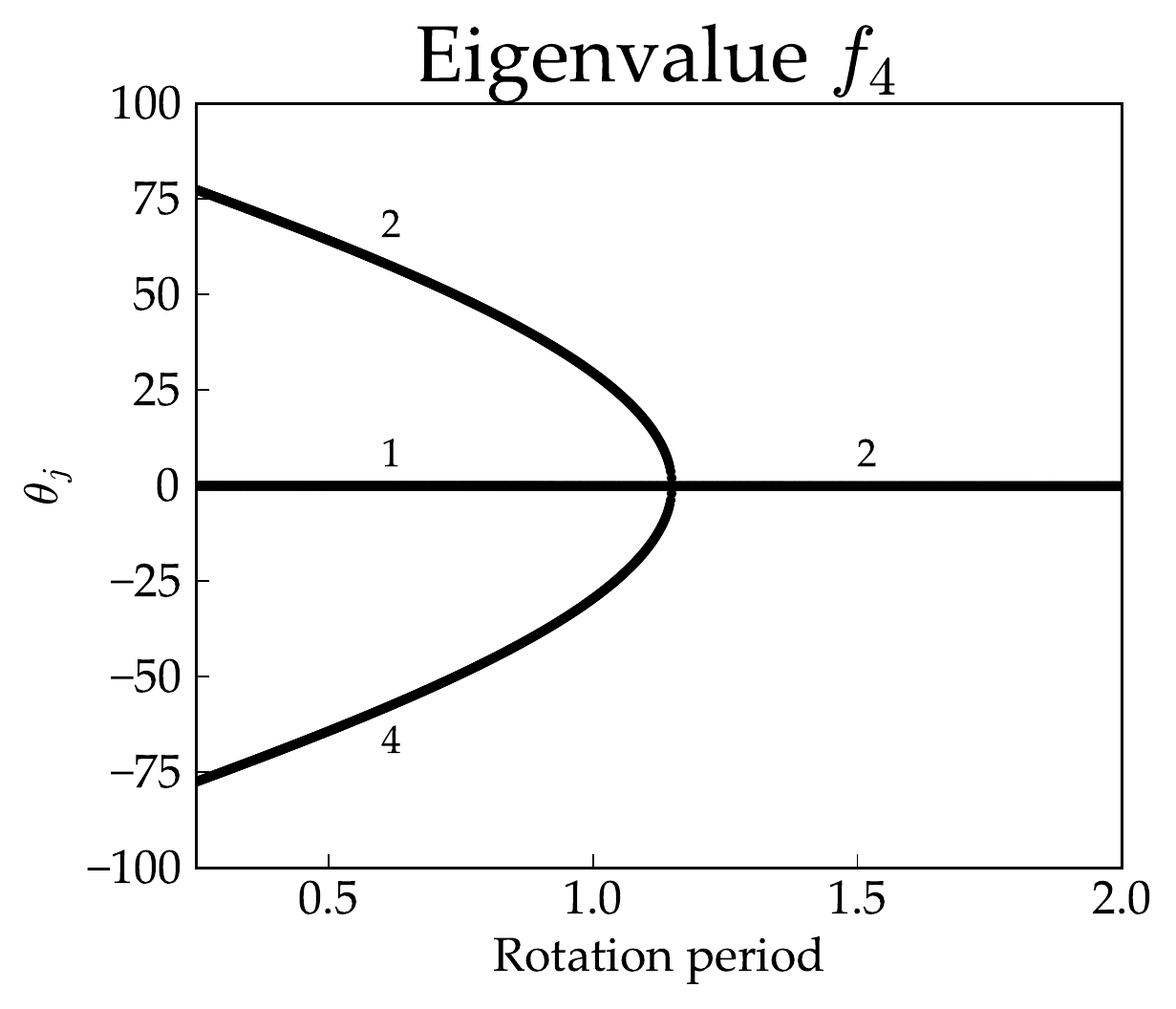} 
\includegraphics[width=0.5\textwidth]{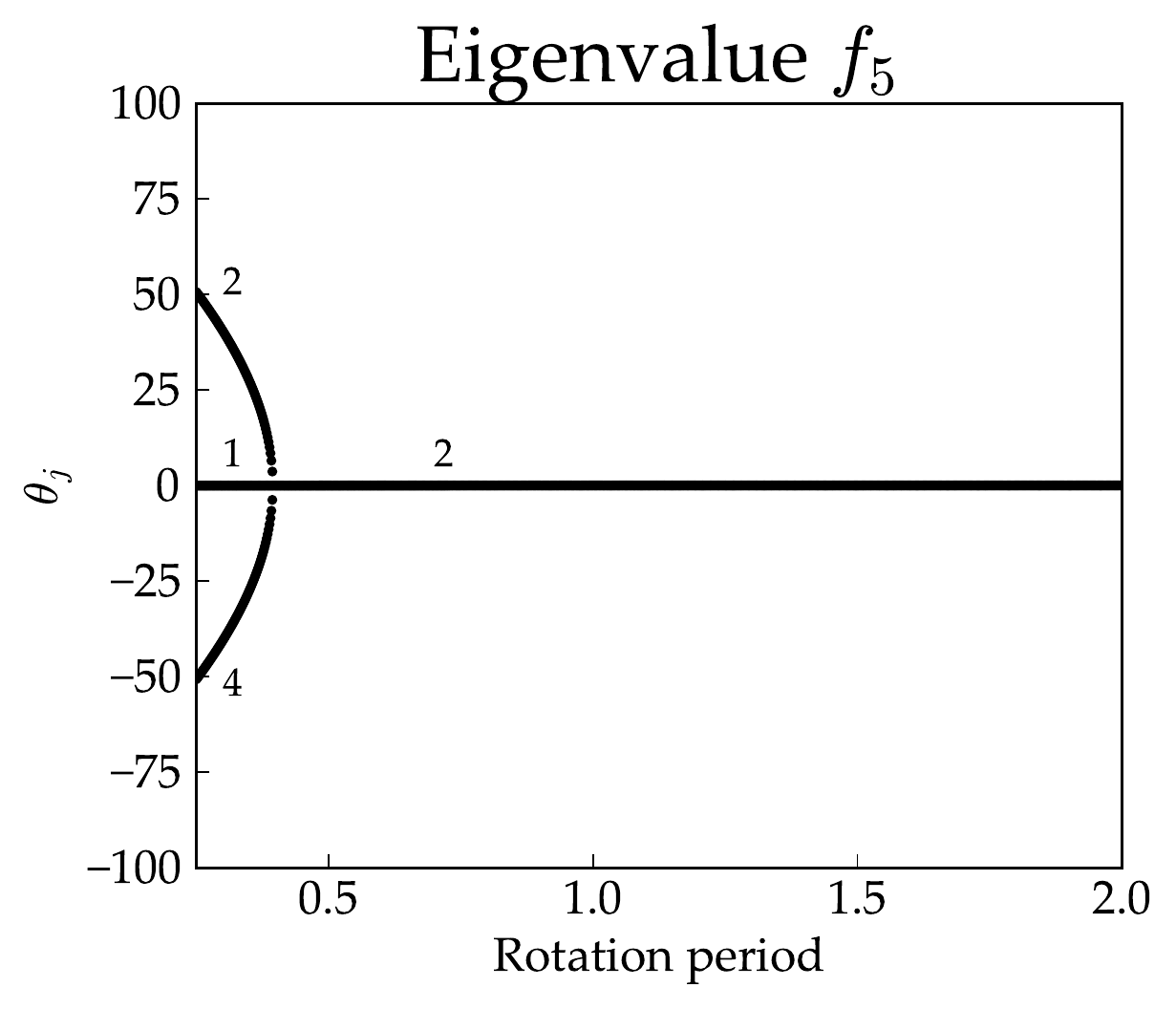}
\caption{\label{k62cassini} Cassini state locations for Kepler-62 f, using our mass set $\mathcal{A}$. The left panel corresponds to the $4$th eigenvalue in inclination (see Table \ref{k62eign1}) and the right to the $5$th eigenvalue. The $4$th eigenvalue has the highest amplitude (see Table \ref{k62amplitude1}) and thus is the strongest forcing term on planet f's inclination and obliquity. At a critical value of $| \alpha/s_i |$, states 1 and 4 merge and then vanish, leaving only states 2 and 3 (not shown)---see \cite{wardhamilton2004}. This occurs for $P_{rot} \approx 1.2$ days for $s_4$ and $P_{rot} \approx 0.35$ days for $s_5$.}
\end{figure*}

From Equations (\ref{eqn:cass13}) and (\ref{eqn:cass24}), we can calculate the locations of Cassini states as a function of rotation period. These are shown in Figure \ref{k62cassini} for the two dominate eigenvalues, $s_4$ and $s_5$. As described in \cite{wardhamilton2004}, Cassini states 1 and 4 merge and vanish at a critical value of $|\alpha/s_i|$ ($\alpha$ depends on $P_{rot}$). Beyond that critical value, only states 2 and 3 exist, where $\theta_2 \approx 0^{\circ}$ and $\theta_3 \approx 180^{\circ}$. Cassini state 3, which is not shown, always has an obliquity of $\approx 180^{\circ}$, corresponding to a retrograde spin.

\begin{table}[h]
\centering
\caption{Inclination amplitudes for each eigenvalue for Kepler-62 f (secular solution)}

\begin{tabular}{lccccc}
\hline\hline \\ [-1.5ex]

& $I_1$  ($^{\circ}$) & $I_2$  ($^{\circ}$)& $I_3$  ($^{\circ}$)& $I_4$  ($^{\circ}$) & $I_5$  ($^{\circ}$) \\
\hline
$\mathcal{A}$ &  2.785$\times 10^{-8}$  & 1.023$\times 10^{-4}$& 2.698$\times 10^{-4}$&  0.1729 & 0.08692 \\
$\mathcal{B}$ & 4.551$\times 10^{-8}$ & 1.025$\times 10^{-4}$& 2.838$\times 10^{-4}$&0.1607 & 0.07788 \\
\hline
\end{tabular}
\label{k62amplitude1}
\end{table}

\begin{figure*}[t]
\includegraphics[width=\textwidth]{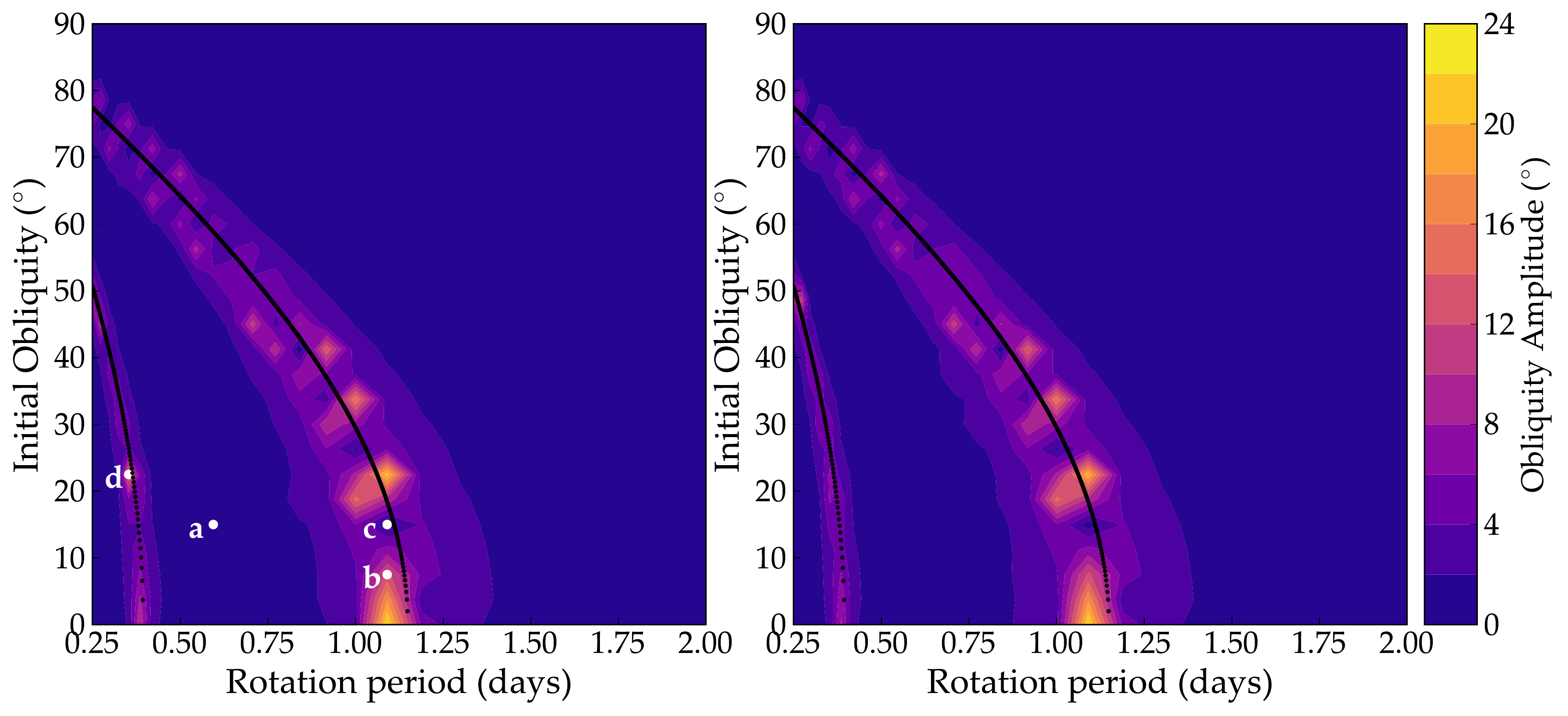}
\caption{\label{k62oblmap} Amplitude of the obliquity oscillation for Kepler-62 f as a function of the initial obliquity and rotation period, for the stable orbital configuration from \cite{bolmont2015} (our mass set $\mathcal{A}$) with initial $\psi + \Omega = 0^{\circ}$. The left panel shows the solution from our secular models \texttt{DISTORB} and \texttt{DISTROT}, and the right panel shows the same solution achieved by coupling \texttt{HNBody} to \texttt{DISTROT}. Secular resonances appear at rotation periods $\lesssim 2$ days. These are minutely shifted in the N-body solution compared to the secular solution (compare the eigenvalues of Tables \ref{k62eign1} and \ref{k62eign2}), because of the small orbital time-scale variations in the semi-major axes of the planets. The secular resonances shown here occur in the proximity of Cassini state 2, for the two eigenvalues $s_4$ and $s_5$. The black curves show the predicted locations of the two Cassini states. The white points correspond to cases shown in Figures \ref{k62orbit1} and \ref{k62orbit2}.}
\end{figure*}

\begin{figure*}[t]
\includegraphics[width=\textwidth]{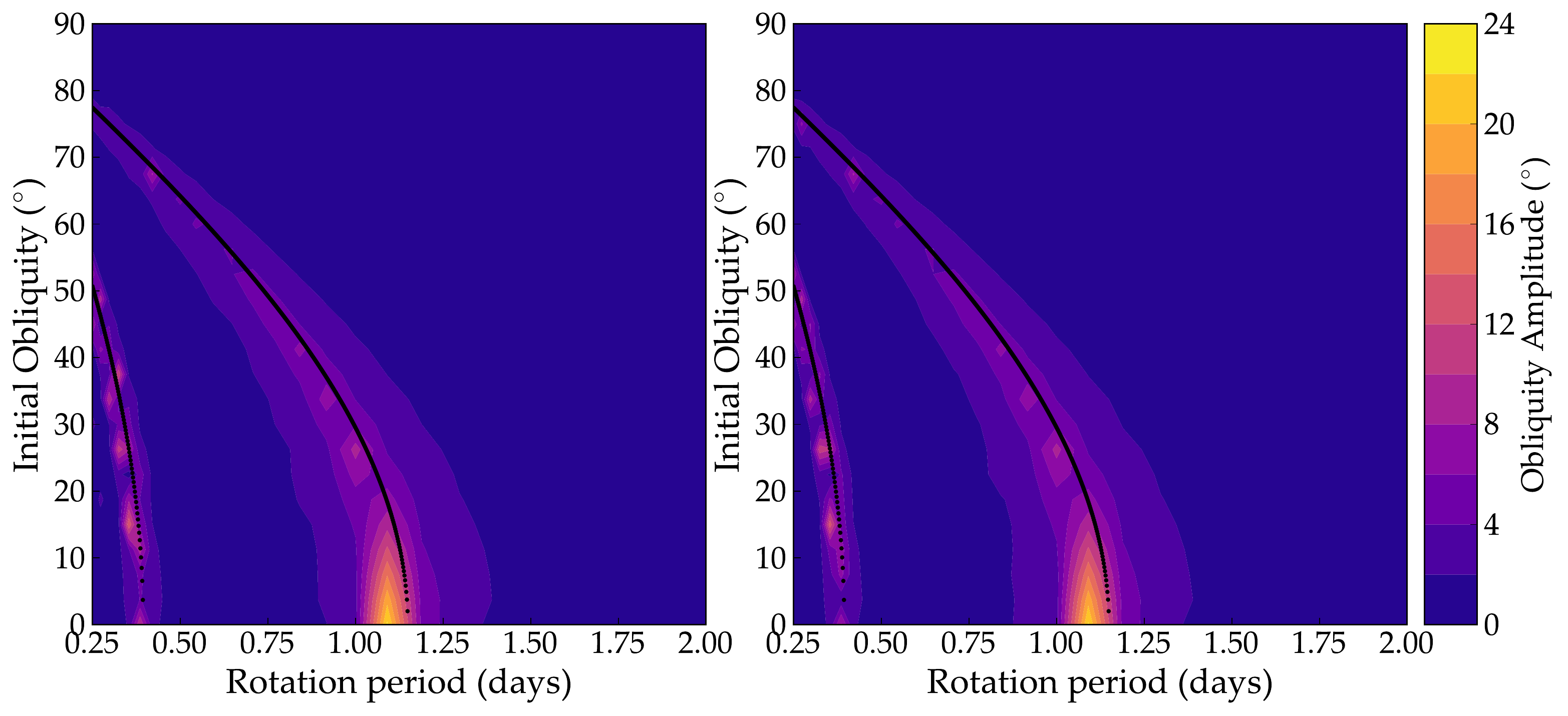}
\caption{\label{k62oblmap2} Same as Figure \ref{k62oblmap} but for initial $\psi + \Omega = 180^{\circ}$. Here, the secular resonances occur in the proximity of Cassini state 4, which is a saddle point.}
\end{figure*}

Figure \ref{k62oblmap} shows the amplitude of the obliquity oscillation for Kepler-62 f, as a function of its initial obliquity and its rotation period, with initial $\psi + \Omega = 0^{\circ}$. The left panel is our secular orbital solution, the right is the N-Body solution (coupling \texttt{HNBody} \citep{rauch2002} to \texttt{DISTROT}). There are two clear secular spin-orbit resonances in this parameter space, each visible as a bright arc with shape $\sim \cos{\varepsilon}$. The black curves show the obliquities associated with Cassini state 2 for eigenvalues $s_4$ and $s_5$. These look very similar to the secular spin-orbit resonance identified for the planet HD 40307 g in \cite{brasser2014}, which we discuss in Section \ref{cassini}. Figure \ref{k62oblmap2} shows the same quantities, but for $\psi + \Omega = 180^{\circ}$. The black curves correspond to the Cassini state 4 obliquities. The secular resonances appear in the same locations here, but the structure is somewhat different from those in Figure \ref{k62oblmap}.

\begin{figure*}[t]
\includegraphics[width=\textwidth]{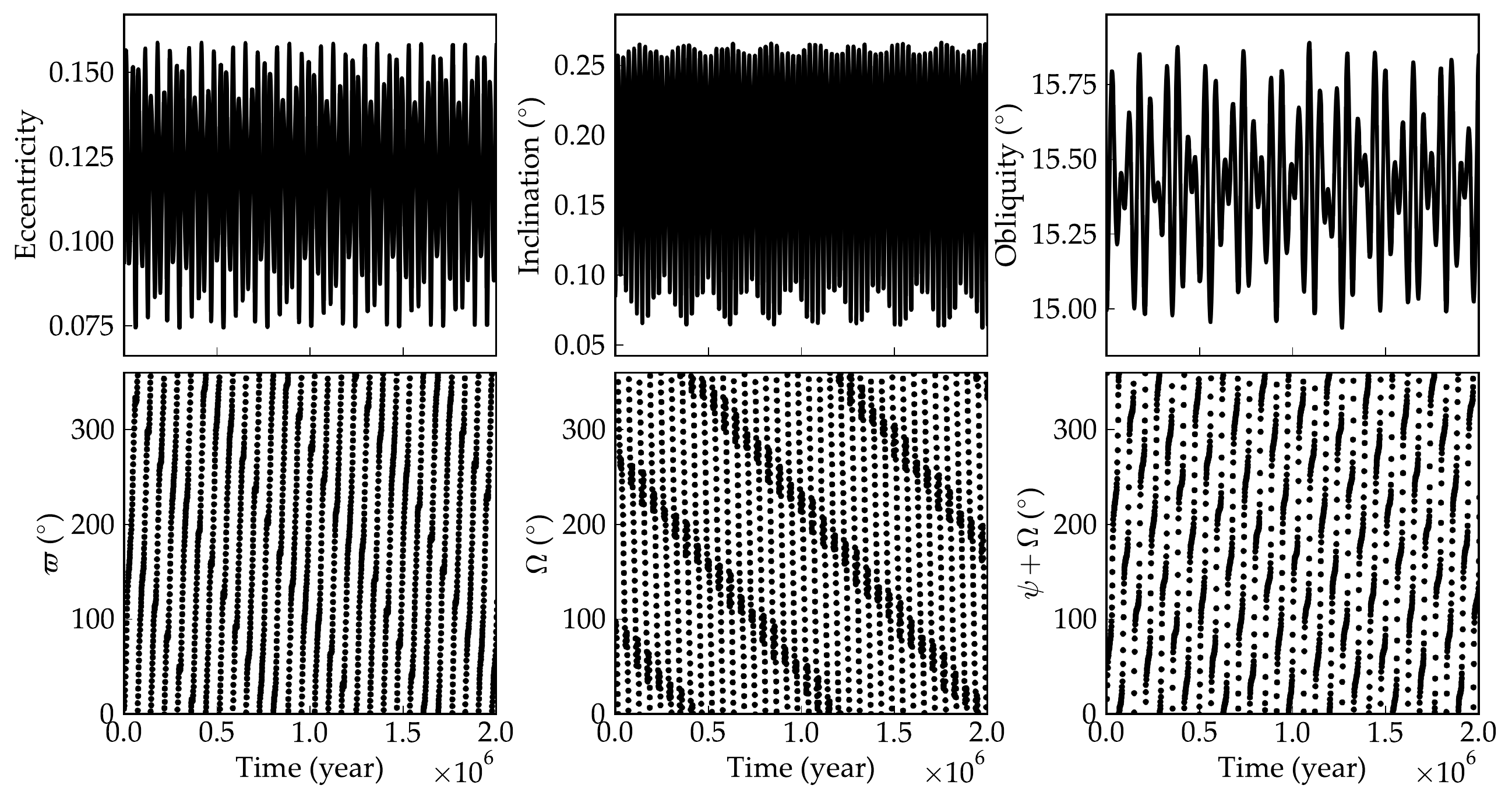}
\caption{\label{k62orbit1} Orbital and obliquity evolution for Kepler-62 f at point {\bf a} in Figure \ref{k62oblmap}, with $P_{rot} = 0.595$ days and $\varepsilon_0 = 15^{\circ}$. The quantity $\psi + \Omega$ is plotted rather than $\psi$, because this angle is diagnostic of secular resonances and Cassini states. Here, the planet is well away from resonance and $\psi+\Omega$ circulates.}
\end{figure*}

\begin{figure*}[t]
\includegraphics[width=\textwidth]{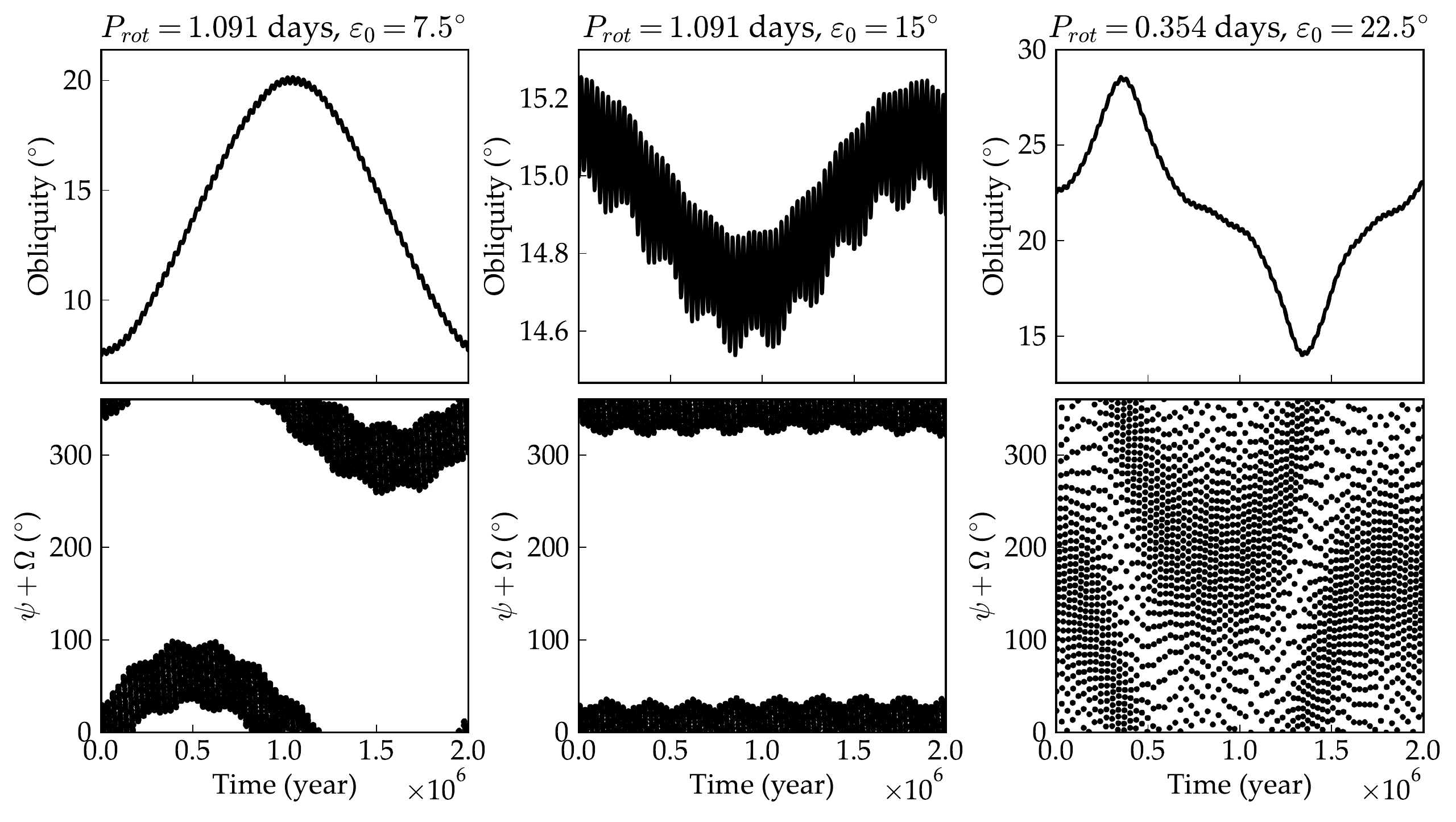}
\caption{\label{k62orbit2} Obliquity evolution at points {\bf b} (left; $P_{rot} = 1.091$ days, $\varepsilon_0 = 7.5^{\circ}$), {\bf c} (middle; $P_{rot} = 1.091$ days, $\varepsilon_0 = 15^{\circ}$), and {\bf d} (right; $P_{rot} = 0.354$ days, $\varepsilon_0 = 22.5^{\circ}$) in the left panel of Figure \ref{k62oblmap}. The orbital evolution is identical to case {\bf a} in Figure \ref{k62orbit1}. At point {\bf b}, the planet is in a secular spin-orbit resonance that pumps its obliquity to $\approx 20^{\circ}$, and $\psi + \Omega$ librates about $0^{\circ}$. It is close to Cassini state 2: the secular resonance can be described as the motion about state 2. At point {\bf c}, the planet is extremely close to Cassini state 2---the argument $\psi+\Omega$ librates more closely about $0^{\circ}$ and the obliquity oscillation becomes much smaller, nearly satisfying 2 of Cassini's laws. At point {\bf d}, the planet is in a spin-orbit resonance with eigenvalue $s_5$. The resonance is not clearly imprinted on the angle $\psi+\Omega$, however, because the eigenvalue $s_4$ still dominates the evolution of $\Omega$.}
\end{figure*}

Figure \ref{k62orbit1} shows the orbital and obliquity evolution for planet f at point {\bf a} in Figure \ref{k62oblmap}. Here, the oscillation of the obliquity is very small and the angle $\psi+\Omega$ circulates. In Figure \ref{k62orbit2}, we show the evolution of the obliquity and $\psi+\Omega$ at points {\bf b}, {\bf c}, and {\bf d}. Both points {\bf b} and {\bf c} are inside the secular resonance associated with frequency $s_4$, and $\psi + \Omega$ librates about $0^{\circ}$; however, the obliquity evolution is markedly different. Point {\bf c} appears to be in or very near to Cassini state 2; thus the $\psi+\Omega$ librates more tightly about $0^{\circ}$ and the obliquity oscillates by only $\sim 0.7^{\circ}$. At point {\bf b}, the obliquity is further from the Cassini state, and oscillates about it with much large amplitude. 

At point {\bf d}, we also see a large obliquity oscillation. The secular resonance here is associated with the secondary frequency $s_5$. This resonance is still able to drive the obliquity strongly, though the angle $\psi +\Omega$ does not clearly librate. This is because $s_4$ is still the stronger driver of the inclination. 

\begin{figure*}[t]
\includegraphics[width=0.5\textwidth]{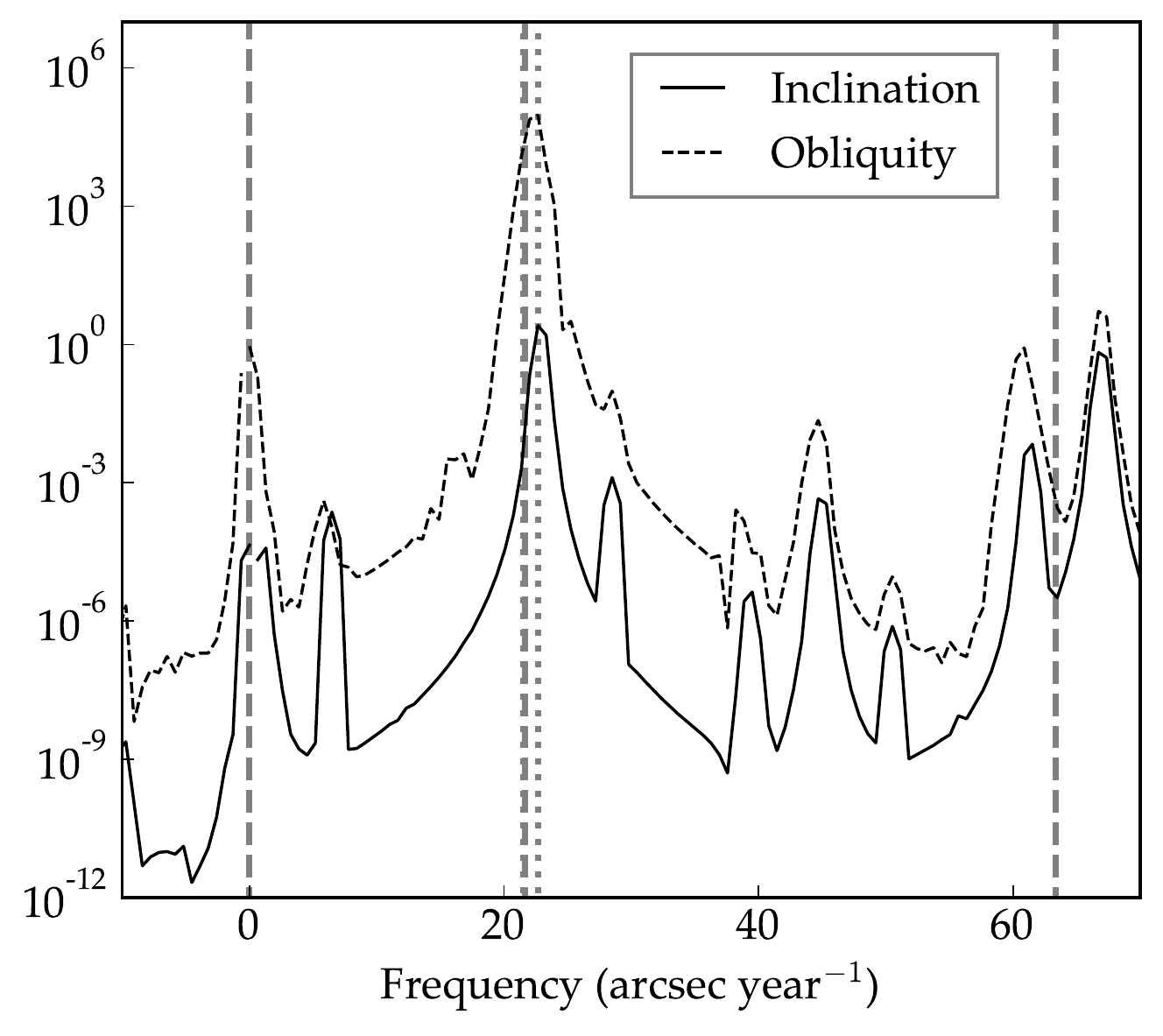}
\includegraphics[width=0.5\textwidth]{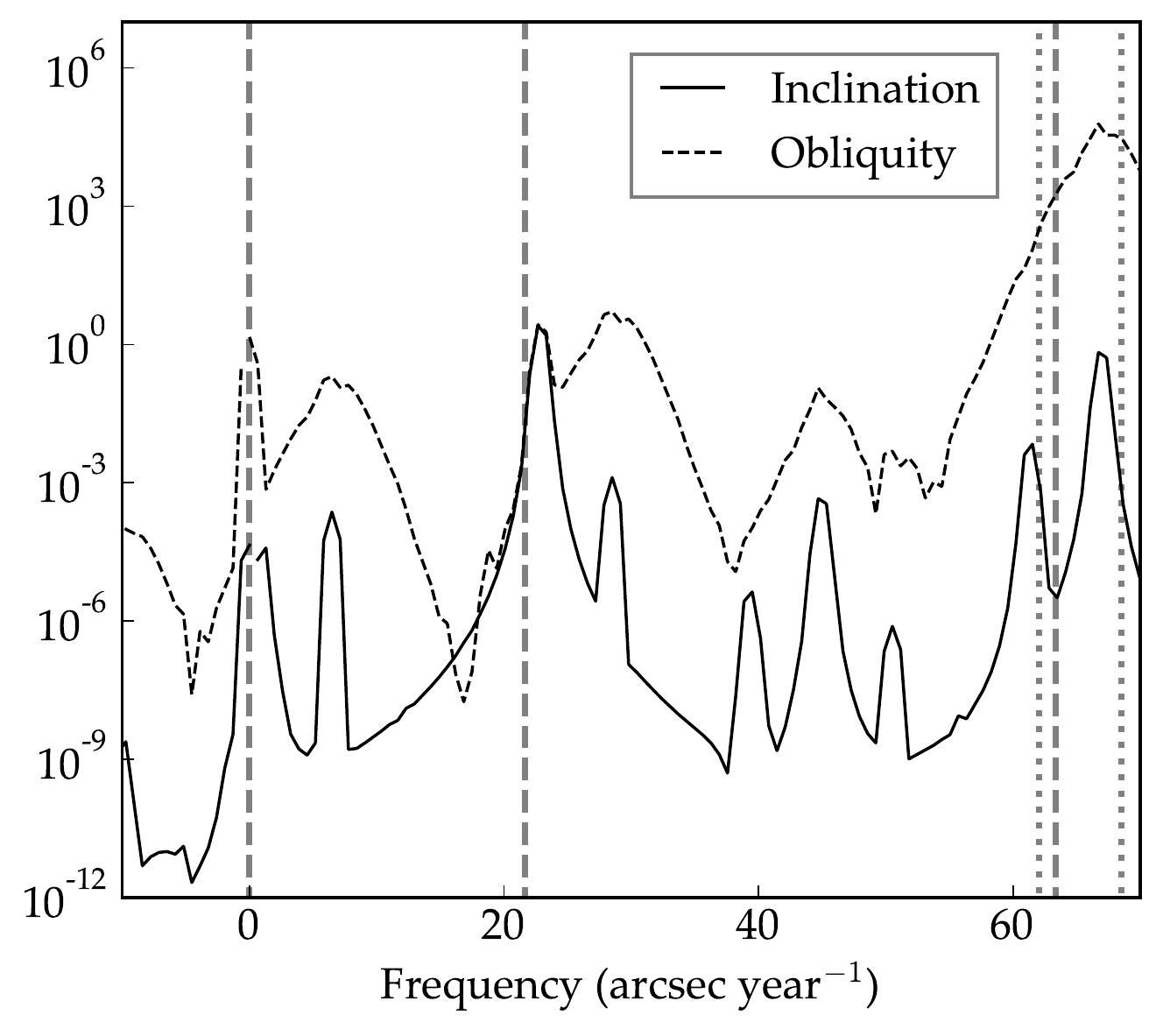}
\caption{\label{k62resfreq1} Power spectra for the obliquity and inclination evolution (left) for Kepler-62 f, in the secular resonances in Fig \ref{k62oblmap}. The left panel corresponds to point {\bf d}, the right to point {\bf b}. The vertical dotted lines are the natural axial precession frequencies from Equation (\ref{eqnR}) at the minimum and maximum obliquities and the vertical dashed lines are the inclination eigenvalues from the Laplace-Lagrange solution. In the right panel, the natural precession rate fall nearly on top the largest inclination frequency, $s_4$. In the left panel, the precession rate lies close to a secondary cluster of frequencies associated with $s_5$. Despite the fact that the argument $\psi+\Omega$ is not seen to librate in this case (Figure \ref{k62orbit2}, right panels), the power spectrum reveals the secular resonance.}
\end{figure*}

Figure \ref{k62resfreq1} presents another way of viewing secular resonances. The left panel shows power spectra of the obliquity variables, $\zeta + \sqrt{-1}\xi$, and the inclination variables, $q+\sqrt{-1}p$, at points {\bf b} (left) and {\bf d} (right). Note that we have inverted the sign of the frequencies in the inclination variables, because the resonance occurs when the frequencies in obliquity and inclination are opposite in sign. This can be seen by rewriting Equations (\ref{eqnpA}) and (\ref{eqnpsi}) in terms of $i$ and $\Omega$ instead of $p$ and $q$ \citep[see][]{laskar1986}. The longitude of ascending node, $\Omega$, and the precession angle, $\psi$ appear together inside sine and cosine functions as $\Omega + \psi$.

The dotted lines in this figure, representing the minimum and maximum values of $R(\varepsilon)$ lie close to peaks in the inclination. In the case at point {\bf b}, the axial precession frequency is very close to the primary inclination frequency, $s_4$, while at point {\bf d}, it is closer to a cluster of peaks at $s_5$. So we see that there is indeed a secular resonance at point {\bf d}, even though the angle $\psi+\Omega$ does not librate.

Regarding the difference in the structure of the $s_4$ secular resonance between Figures \ref{k62oblmap} and \ref{k62oblmap2}: the initial angle $\psi +\Omega$ places the planet near Cassini state 2 in Figure \ref{k62oblmap} and near Cassini state 4 in Figure \ref{k62oblmap2}. Since state 2 is stable, there are ``islands'' of small obliquity oscillation very close to the black curve in Figure \ref{k62oblmap}. Since state 4 is a saddle point, such islands do not appear in Figure \ref{k62oblmap2}; the ``lumpiness'' in the structure of the resonance is a result of the discrete nature of the grid of simulations and the interpolation used in making the contours.

\begin{figure*}[t]
\includegraphics[width=\textwidth]{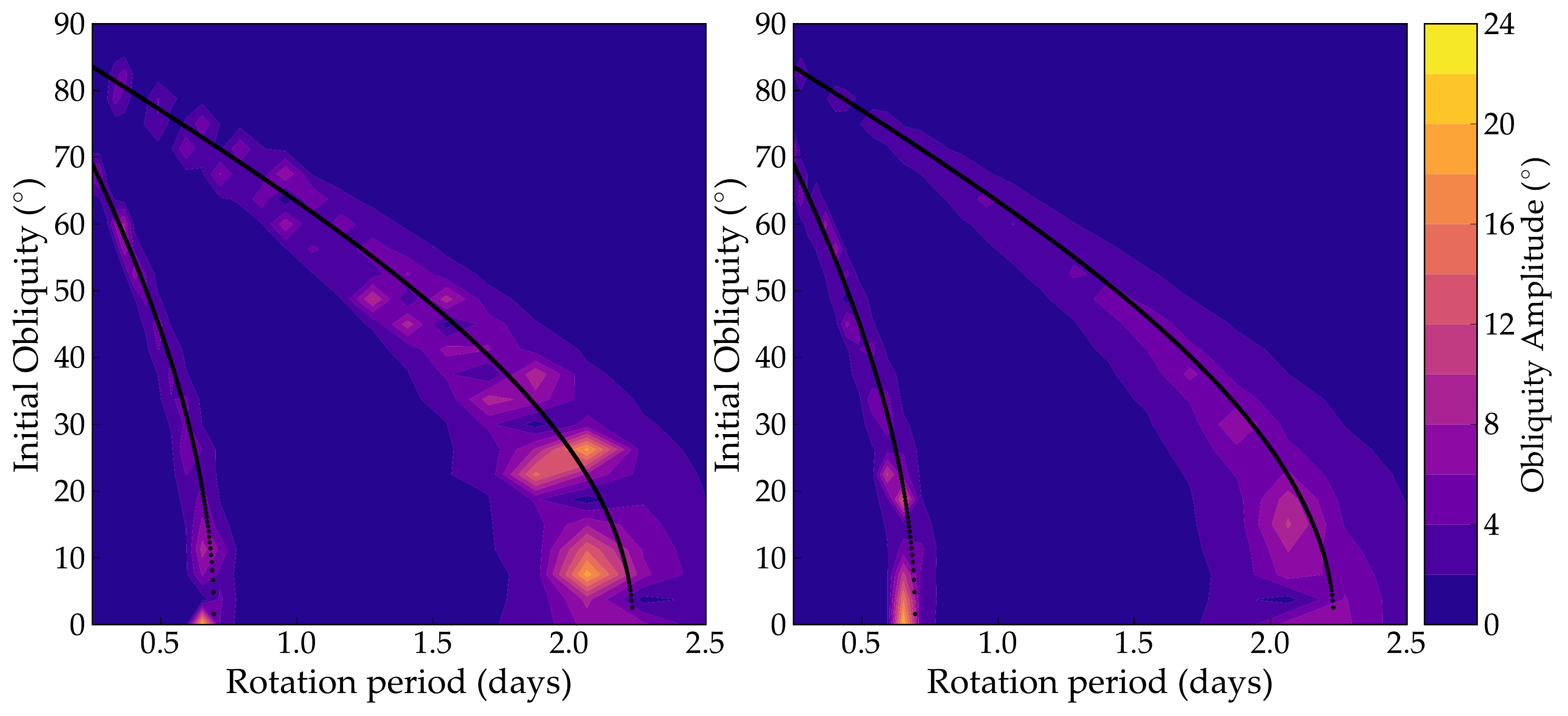}
\caption{\label{k62oblmap3} Same as Figure \ref{k62oblmap} but using mass set $\mathcal{B}$. The left panel has $(\psi+\Omega)_0  = 0^{\circ}$, the right $(\psi+\Omega)_0  = 180^{\circ}$. The secular resonances (and Cassini states) are shifted to longer rotation periods.}
\end{figure*}

Finally, Figure \ref{k62oblmap3} shows the obliquity amplitude for mass set $\mathcal{B}$. The smaller masses of the planets in this case lower all of the frequencies, as compared with mass set $\mathcal{A}$ (see Tables \ref{k62eign1} and \ref{k62eign2}). Further, the $J_2$ value of planet f is increased (see Equation \ref{eqnJ2}), which increases its axial precession rate. The combination of the two effects pushes the resonances toward longer periods.

\subsection{Comparison with HD 40307 g}
\label{cassini}
In their dynamical analysis of the HD 40307 system, \cite{brasser2014} noticed the existence of a secular resonance that can amplify the obliquity oscillation of the outermost planet, depending on the planet's initial obliquity and rotation rate, with a similar structure as we find for Kepler-62 f. Note that the existence of planet HD 40307 g is currently under scrutiny \citep{diaz2016}. Even if planet g does not exist, the system as modeled by \cite{brasser2014} is nonetheless interesting as a test case of dynamical phenomena. 

\begin{figure*}[t]
\includegraphics[width=\textwidth]{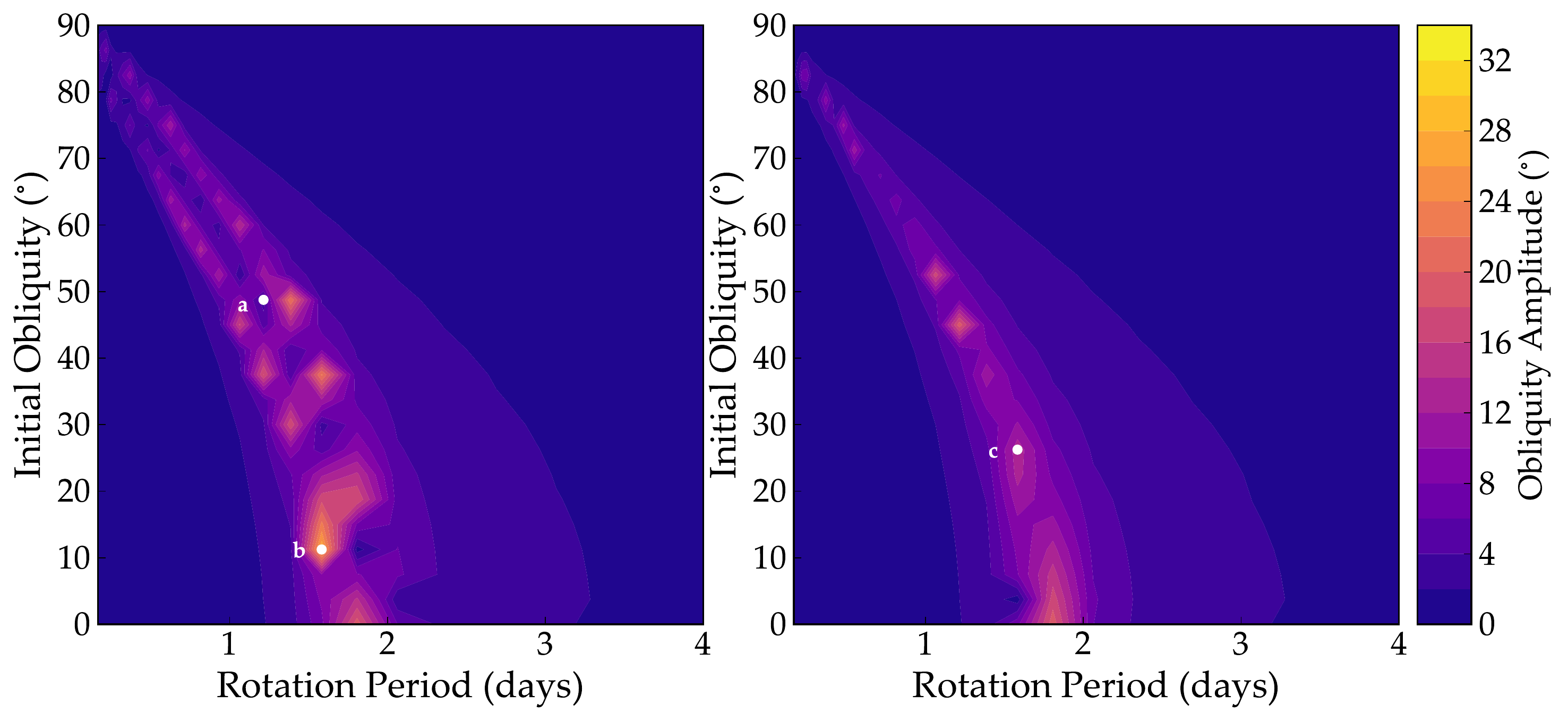}
\caption{\label{hd403_1} Amplitude of the obliquity oscillation for HD 40307 g, as a function of initial obliquity and rotation period. Compare to Fig. 8 in \cite{brasser2014}. The left panel started with $\psi+\Omega=0^{\circ}$, the right with $\psi+\Omega=180^{\circ}$. There is a secular resonance at rotation rates $\lesssim 2$ days, similar to resonances for Kepler-62 f in Fig. \ref{k62oblmap}.}
\end{figure*}

We have simulated HD 40307 system in our model according to the initial conditions used by \cite{brasser2014}, with a starting $\psi = 0^{\circ}$. Plotting the amplitude of the obliquity oscillation of planet g as a function of its initial rotation period and obliquity, we roughly reproduce their Figure 8 (upper left panel) in our Figure \ref{hd403_1} (we adjusted the $x$-axis to better display the resonance). The left panel of Figure \ref{hd403_1} shows our results using our 4th order model for $\psi+\Omega=0^{\circ}$, the right for $\psi+\Omega=180^{\circ}$. Including 4th order terms reduces the amplitude of the oscillation (as compared to \cite{brasser2014}, Figure 8) and shifts the resonance to slightly shorter rotation periods. As \cite{brasser2014} explain, this resonance can be easily identified in the eigenvalue solution, just as we showed for the two resonances in Kepler-62. 

\begin{figure*}[t]
\includegraphics[width=\textwidth]{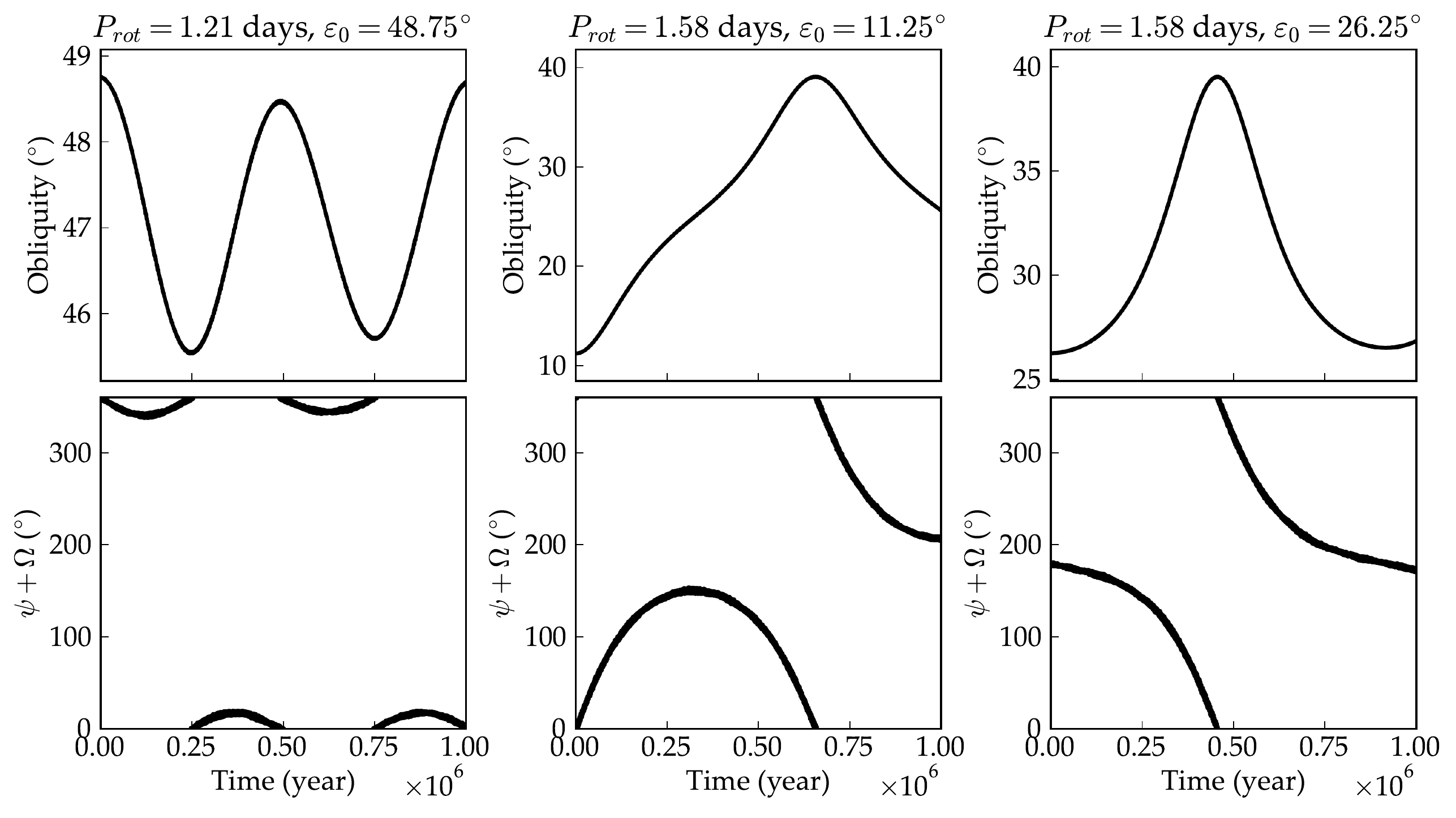}
\caption{\label{hd403_2} Evolution of the obliquity and the angle $\psi+\Omega$ for points {\bf a} (left), {\bf b} (middle), and {\bf c} (right) from Figure \ref{hd403_1}. Point {\bf a} approaches Cassini state 2 wherein the obliquity oscillation is smaller than the surrounding regions. Point {\bf b} is librating about Cassini state 2 with a large amplitude and its obliquity oscillation is large. Point {\bf c} is near Cassini state 4.}
\end{figure*}

It is not clearly explained how those authors determined that planet g (in their simulations) is in a Cassini state inside the secular resonance, where the obliquity oscillates by the largest amount. However, much as we see for Kepler-62 f, there is a relationship between Cassini states 2 and 4 and the secular resonance. Which Cassini state the planet is near depends on the initial value of the angle $\psi+\Omega$. In the left panel of Figure \ref{hd403_1}, $\psi+\Omega = 0$ and the planet is near the stable Cassini state 2. In the right panel, $\psi+\Omega=180$ and the planet is near the unstable state 4. Near each state, the obliquity oscillates by a large amount as the spin axis circulates about the Cassini state obliquities. However, for state 2, there is a point (as a function of rotation rate) at which the obliquity can become approximately fixed, and we see the ``islands'' of low obliquity oscillation inside the resonance.

\subsection{Test system: Earth-mass planet and eccentric giant}
\label{sectsys1}

Our system, TSYS, is simulated as a test case of a longer period Earth-mass planet perturbed by a giant planet and a test of the effects of large mutual inclinations and eccentricity. The purpose of this investigation is to demonstrate the complexity and large-amplitude obliquity variations that are possible in dynamically hot systems, which may be discovered in the coming decades. This is a fictitious case of an Earth-mass planet in the habitable zone of a G-dwarf, with a warm Neptune at $\sim0.13$ au and a Jovian at $\sim 4$ au. Interestingly, the inner-most planet turns out to have a minimal effect on the dynamical evolution of the Earth-mass planet; instead, its orbital and obliquity variations are strongly driven by the Jovian at $\sim4$ au.

\begin{figure}[t]
\includegraphics[width=\textwidth]{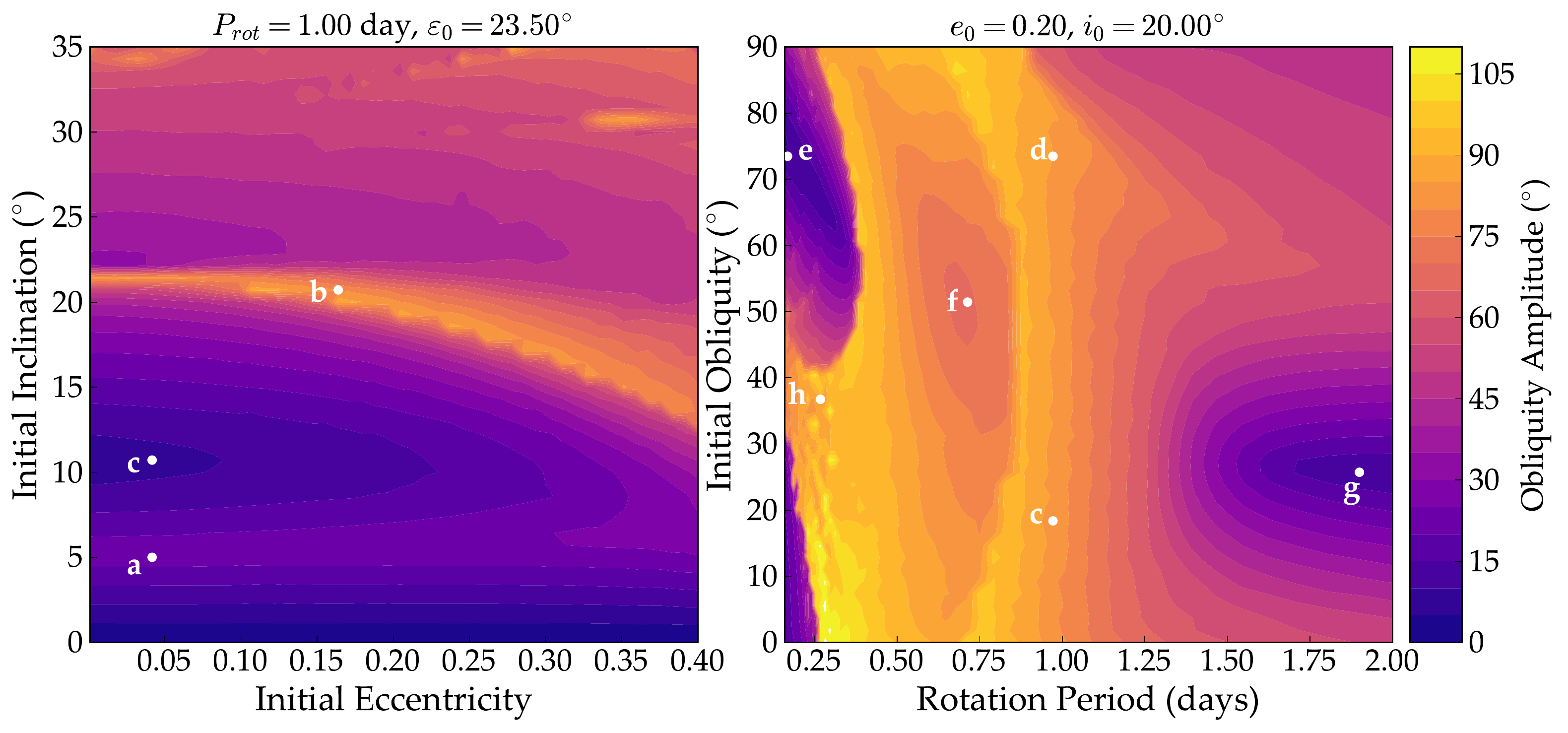}
\caption{\label{sys1oblvsrot} Amplitude of the obliquity oscillation for the Earth-mass planet in TSYS, over two slices of our parameter space. On the left, we vary $e_0$ and $i_0$ at a fixed $P_{rot} = 1$ day and $\varepsilon_0 = 23.5^{\circ}$. A strong secular spin-orbit resonance appears as a curved arc across this space. At the minimum obliquity oscillation on the lower left (point {\bf a}), Cassini's third law is satisfied. On the right, we vary the initial obliquity and rotation period at fixed $e_0 = 0.2$ and $i_0 = 20^{\circ}$. Large amplitude obliquity oscillations appear as bright yellow structures throughout this space. At points {\bf c} and {\bf d}, the motion of the obliquity is very similar to that at point {\bf b} (see text). Minima appear at points {\bf e}, {\bf f}, {\bf g}---``islands'' of low amplitude obliquity oscillations. Power spectra indicate that these Cassini-like states are associated with 2 of the dominant inclination frequencies.}
\end{figure}

The panels of Figure \ref{sys1oblvsrot} show the maximum change in obliquity of the Earth-mass planet over 2 Myr in two slices through parameter space. In general, the amplitude of the obliquity oscillation is expected to be related to the size of the inclination variation, which increases with initial inclination \citep[see][]{ward1992}. We do see this amplitude correlation, particularly with an initial obliquity of $23.5^{\circ}$ (left-hand panels), however, there are other noteworthy regions that have substantially larger obliquity oscillations. In particular, there is a curved feature (the ``arc'') across the center of the left-hand plot in which the obliquity oscillates by $\sim70^{\circ}$ to $\sim110^{\circ}$, and a minimum that appears below the arc in the left hand panel.

\begin{figure*}[t]
\includegraphics[width=\textwidth]{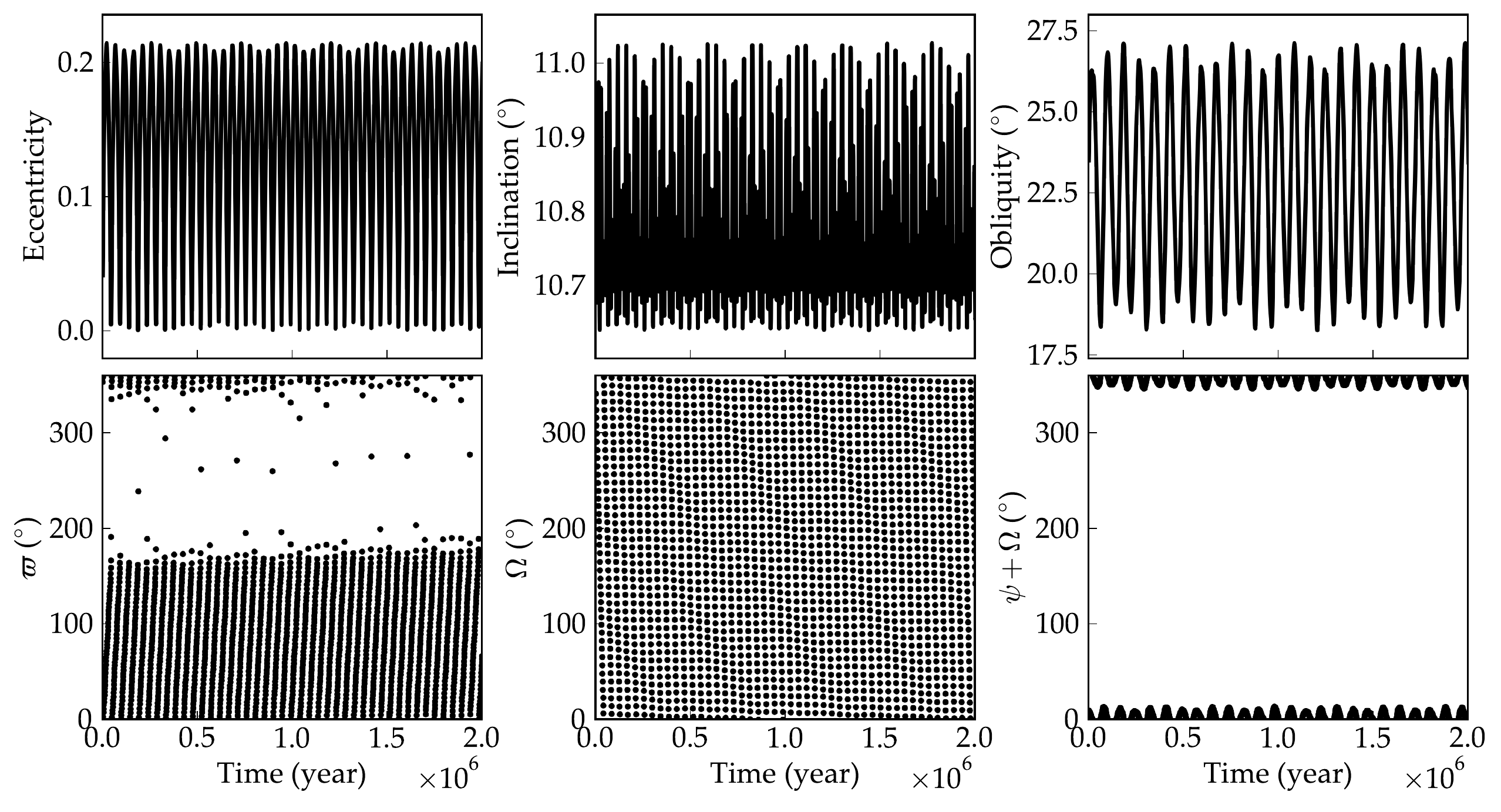}
\caption{\label{orb1sys1} Orbital and obliquity evolution for the Earth-mass planet in TSYS, near a minimum in the obliquity amplitude (point {\bf a} in Fig. \ref{sys1oblvsrot}). Cassini's third law is satisfied ($\psi + \Omega$ librates about $0^{\circ}$), though the obliquity still oscillates by $\sim 10^{\circ}$.}
\end{figure*}

The prominent arc is a result of a secular spin-orbit resonance \citep{ward1992}. The minimum (point {\bf a}) is possibly associated with a Cassini state---we refer to this as a ``Cassini-like'' state, in that Cassini's third law is satisfied, though there does not appear to be a true secular resonance, in which the natural precession rate, $R(\varepsilon)$, would be commensurate with a nodal frequency. Figure \ref{orb1sys1} shows the orbital and obliquity evolution for a case with initial eccentricity $e =0.0417$ and initial inclination $i =10.71^{\circ}$ (point {\bf a} in Figure \ref{sys1oblvsrot}). Here, Cassini's third law is satisfied, though the second is not---the obliquity still oscillates by $\sim 10^{\circ}$.

\begin{figure*}[t]
\includegraphics[width=\textwidth]{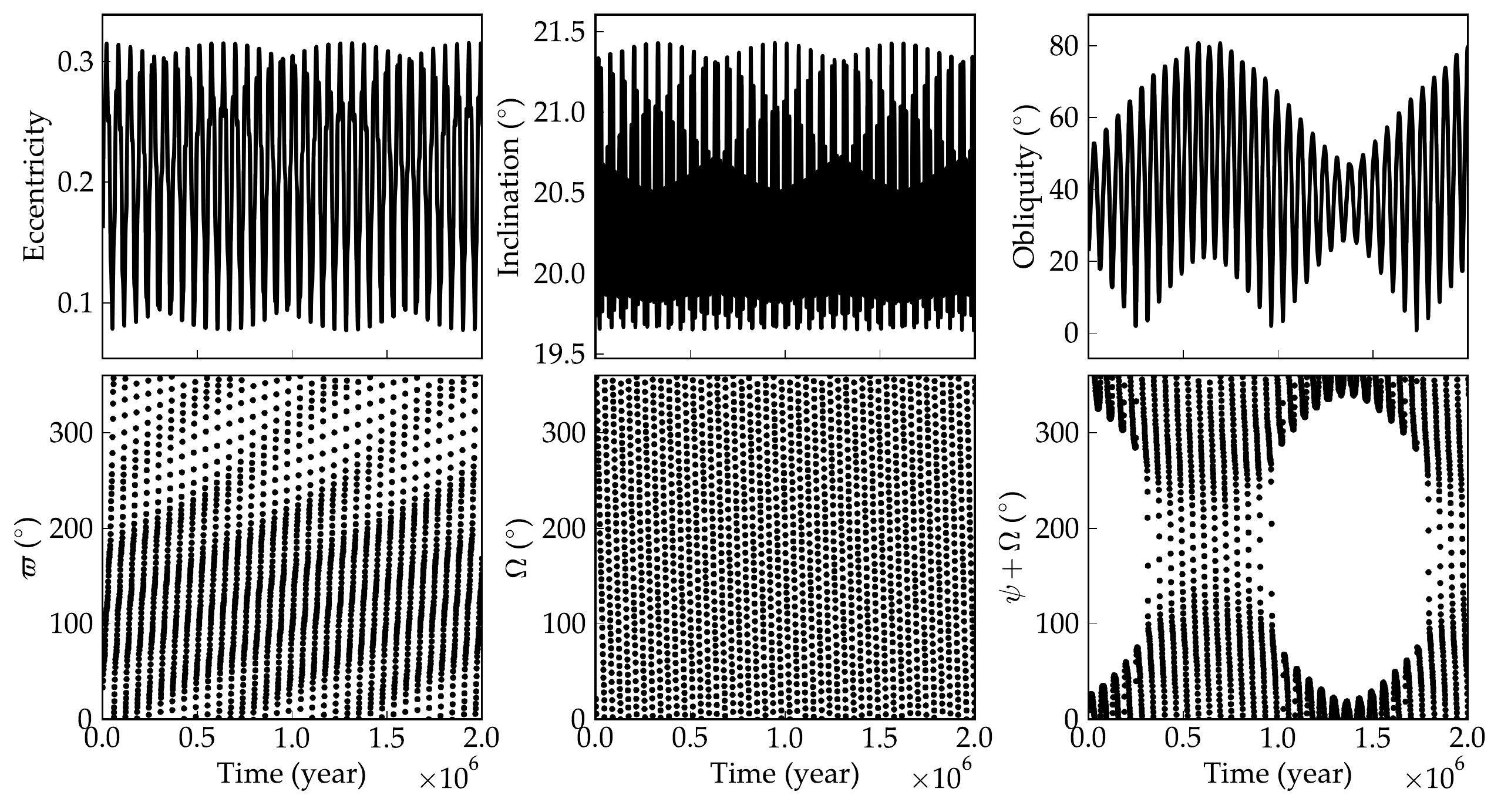}
\caption{\label{orb2sys1} Same as Fig. \ref{orb1sys1}, but for a configuration within the secular resonance (point {\bf b} in Fig. \ref{sys1oblvsrot}). The angle $\psi + \Omega$ librates for much of the time, though not as tightly as for point {\bf a} near a Cassini-like state, and periodically switches to circulation.}
\end{figure*}

The obliquity evolution is particularly dramatic in Figure \ref{orb2sys1}, within the secular resonance. We see that over $\sim1.5$ Myrs, the planet experiences obliquities from nearly zero to $80^{\circ}$. Such variations, along with the somewhat large oscillations in eccentricity, would have a large influence on the climate of a planet with an atmosphere similar to Earth's.  An interesting component to this evolution is that the obliquity and eccentricity evolve with nearly the same frequency, in sync with each other. It would seem then that the variation in $e$ is tightly coupled to the variation in $i$, which drives the obliquity, and hence all three evolve commensurately. 

\begin{figure*}[t]
\includegraphics[width=0.5\textwidth]{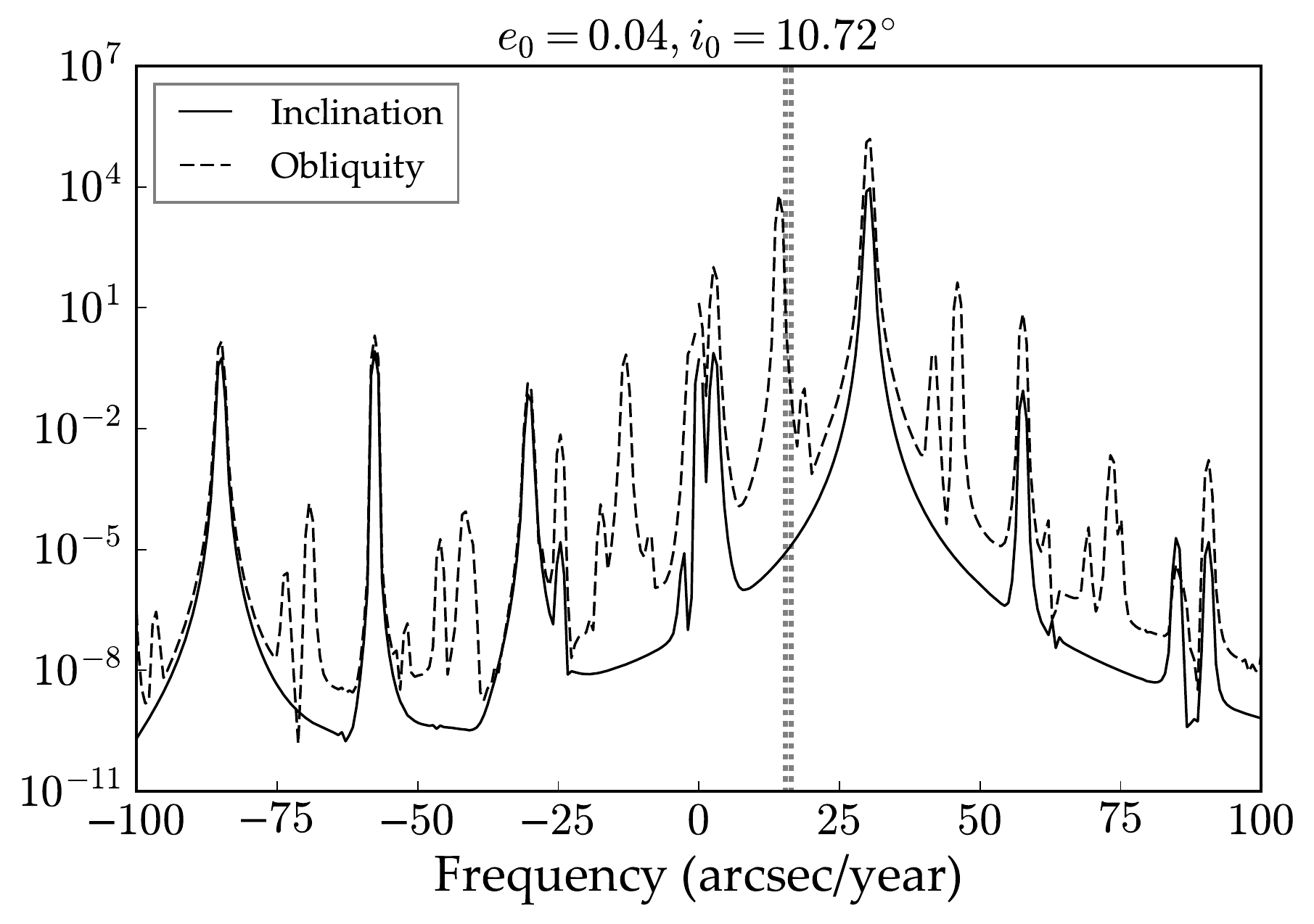}
\includegraphics[width=0.5\textwidth]{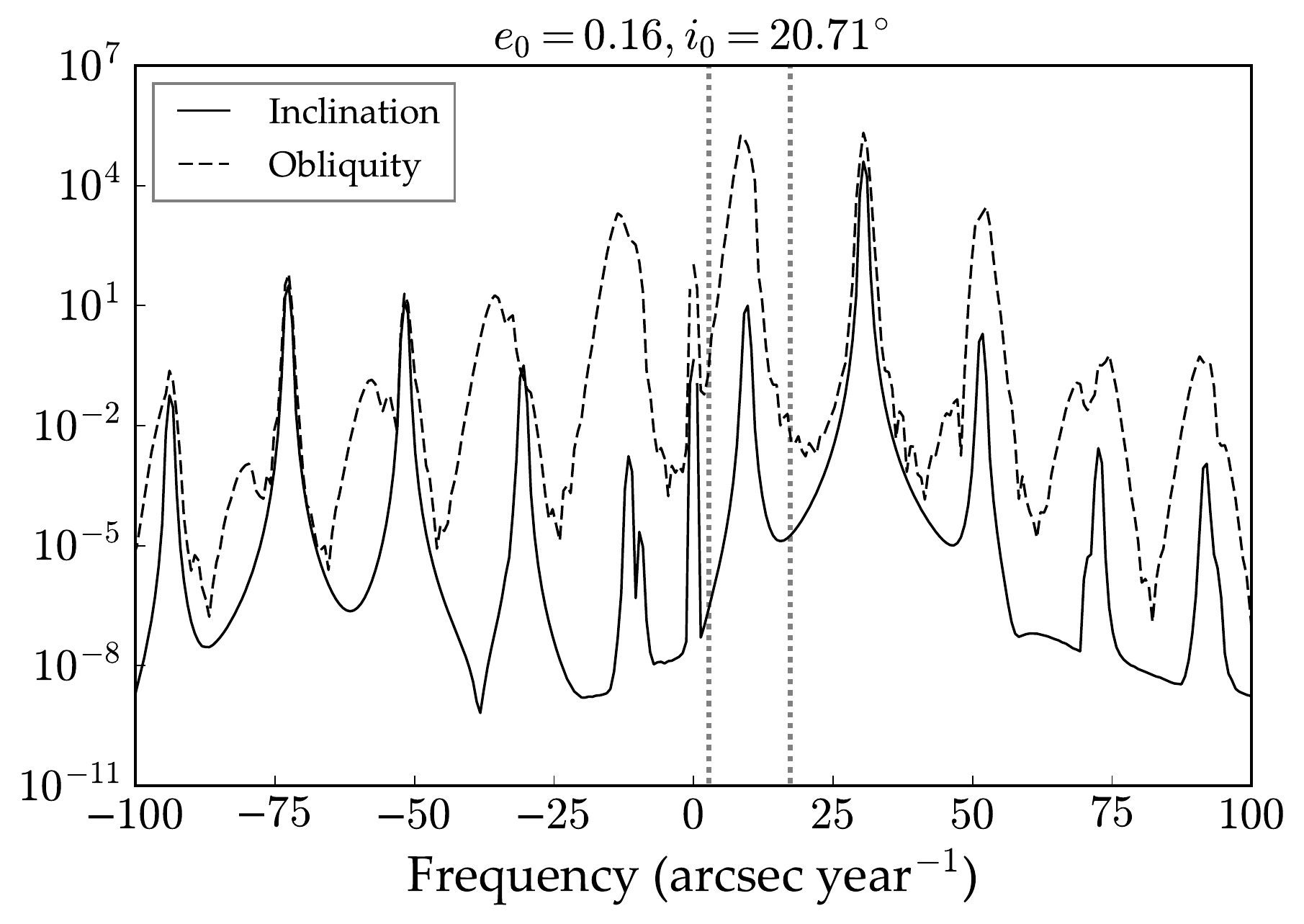}
\caption{\label{freqssys1} Power spectra of the obliquity and inclination evolution for the Earth-mass planet in TSYS, at points {\bf a} (left) and {\bf b} (right) from Figure \ref{sys1oblvsrot}. The vertical dotted lines correspond to the minimum and maximum \emph{natural} axial precession rate (Equation (\ref{eqnR})), which varies because the obliquity varies. Outside the secular resonance, the axial precession frequency fall in between peaks in the inclination spectrum. Within the resonance, this natural precession frequency falls right on top of an inclination peak at $\approx$ 10 arcsec year$^{-1}$.}
\end{figure*}

Figure \ref{freqssys1} shows power spectra for both of the cases. The vertical dashed lines represent the natural precession frequency as calculated from Equation (\ref{eqnR}) at the minimum and maximum obliquity over the 2 Myr simulation. The curves show the power calculated from the obliquity variables, $\zeta + \sqrt{-1}\xi$, and the inclination variables, $q+\sqrt{-1}p$. In both panels, the dominant inclination peaks appear in the obliquity power spectrum because they drive the obliquity, as explained above. We can see in the case on the left that the natural precession rate (vertical dashed lines) falls in a valley between several strong inclination peaks, indicating that this case is not in a secular spin-orbit resonance. Interestingly, the inclination forcing of the obliquity is stronger than the natural precession rate. The libration of $\psi+\Omega$ is a natural consequence of  the inclination cycle dominating over the stellar torque.
This case is located at a minimum in the obliquity oscillation, suggesting that Cassini's second law could be satisfied with a slightly different set of initial conditions.

In the case shown in the right hand panel of Figure \ref{freqssys1}, the natural precession rate falls directly on top of one of the inclination peaks and is a true secular spin-orbit resonance. Refering back to Figure \ref{orb2sys1}, we can see that the obliquity response is indeed very strong as a result of the resonance, evolving from a relatively small oscillation of about $\sim40^{\circ}$ to an almost $80^{\circ}$ swing.

We also tested different starting precession angles in 90$^{\circ}$ intervals, to test this parameter's role, still varying $e_0$ and $i_0$. The secular resonance in this parameter space is insensitive to $\psi$, and appears in all orientations of the equinox ($\psi = 11.78^{\circ}, 101.78^{\circ}, 191.78^{\circ}, 281.78^{\circ}$). However, the minimum in Figure \ref{sys1oblvsrot} (in which point {\bf a} is located) disappears for $\psi \neq 281.78^{\circ}$. This minimum is purely a result of the initial conditions; nothing drives the system to this state.

Because of the large mutual inclination and eccentricities we explore in this system, the resonance and Cassini states are not easily explained in terms of the inclination eigenvalues from the Laplace-Lagrange solution---for example, since the eigenvalues do not depend on eccentricity or inclination, they are the same at every point within the left panel of Figure \ref{sys1oblvsrot}. Yet, the secular resonance clearly depends on both orbital elements. Further, the secular resonance does not follow the $\cos{\varepsilon}$ shape (right panel of Figure \ref{sys1oblvsrot}) that we expect from Equation (\ref{eqnR}) and that we see for Kepler-62 f and HD 40307 g. The origin of this feature is unclear, but is probably due to the complex non-linear dynamics of systems with large mutual inclinations.

\begin{figure*}[t]
\includegraphics[width=0.5\textwidth]{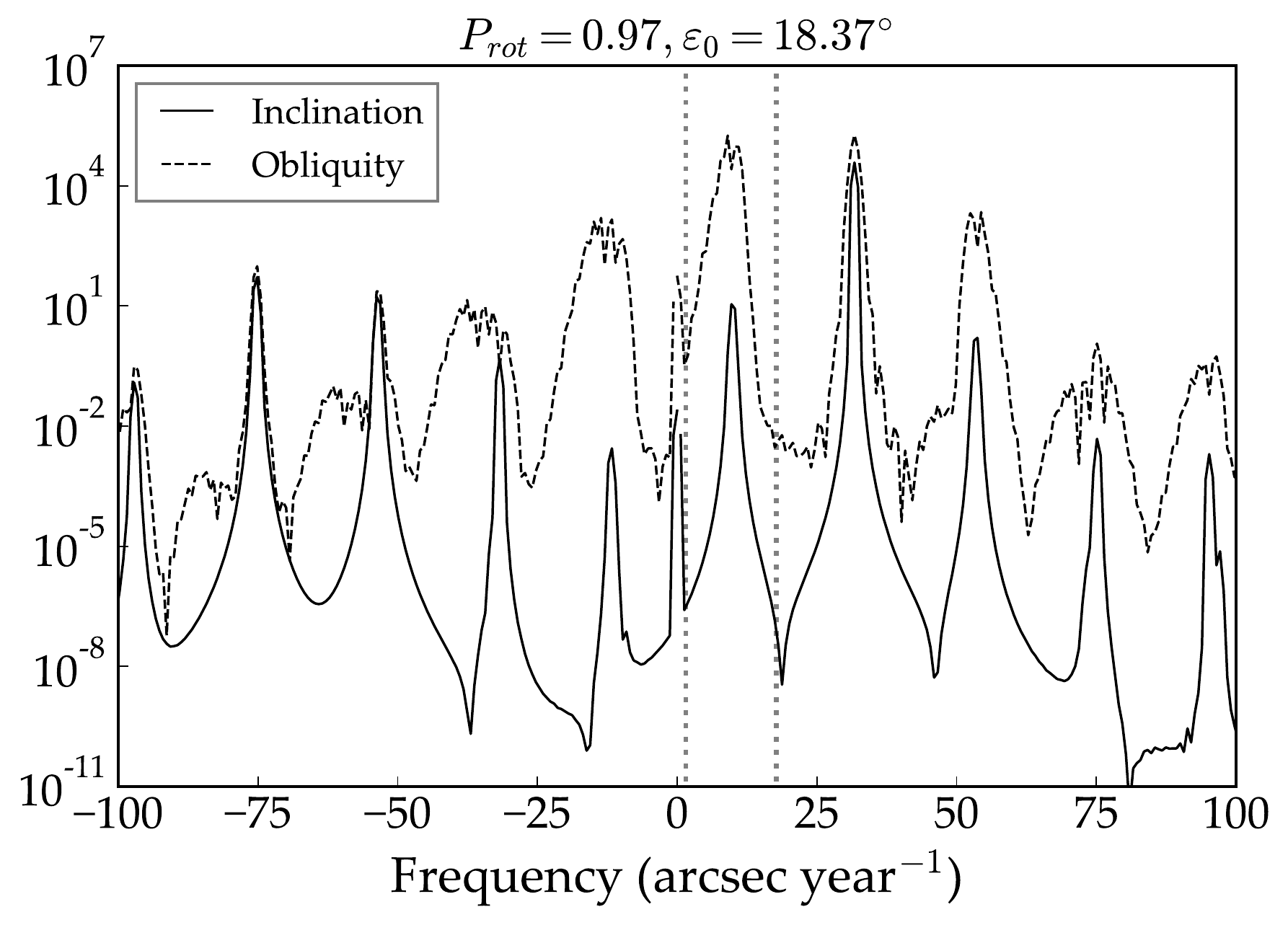}
\includegraphics[width=0.5\textwidth]{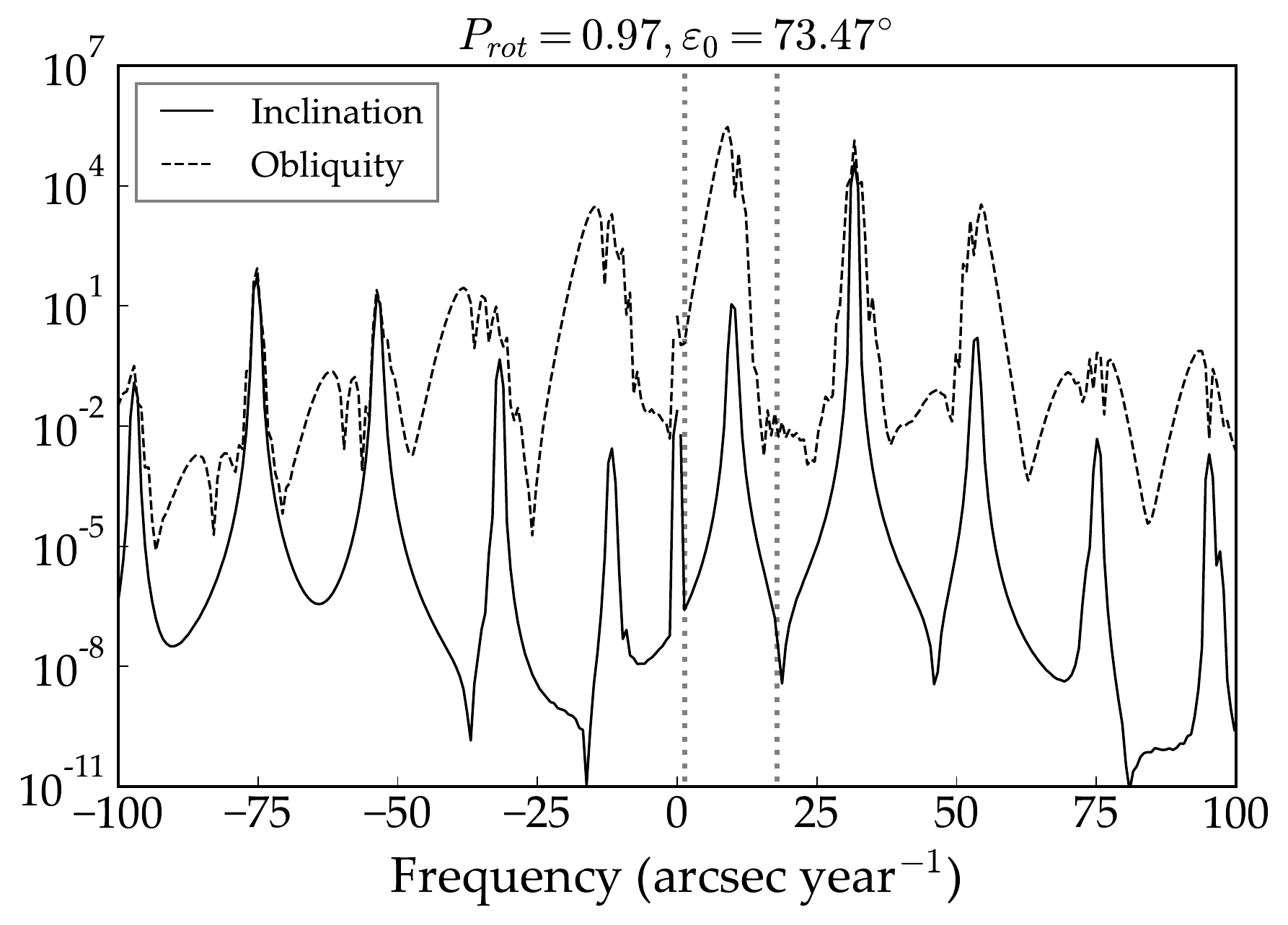}
\caption{\label{freqs2sys1} Power spectra of the obliquity and inclination evolution for the Earth-mass planet in TSYS, at points {\bf c} (left) and {\bf d} (right) from Figure \ref{sys1oblvsrot}, in the same format as Figure \ref{freqssys1}.}
\end{figure*}

Figure \ref{freqs2sys1} shows the power spectra of the obliquity and inclination variables for points {\bf c} and {\bf d} from Figure \ref{sys1oblvsrot}. The two obliquity spectra are nearly identical, despite a difference in initial obliquity of $55^{\circ}$. In fact, the time evolution of the obliquity in both cases looks extremely similar to that of case {\bf b} (Figure \ref{orb2sys1}), only shifted in phase. The two cases have very different initial precession rates ($\sim16.9$ arcsec year$^{-1}$ at point {\bf c} and $\sim5.1$ arcsec year$^{-1}$ at point {\bf d}), but the obliquity oscillation is so large that they end up exploring the same range of frequencies. Thus both cases cross the nodal/inclination peak at $\sim 9$ arcsec year$^{-1}$, passing through resonance, and driving still larger obliquity oscillations.

At points {\bf e}, {\bf f}, and {\bf g}, the obliquity amplitude is reduced compared to the surrounding regions. Each of these ``islands'' is associated with one of the major inclination frequencies: at point {\bf e}, the natural axial precession rate is close to the strongest inclination peak at $\sim 31$ arcsec year$^{-1}$. At points {\bf f} and {\bf g}, the precession is close to the peak at $\sim 9$ arcsec year$^{-1}$. This suggests, again, behavior akin to Cassini states, though involving eccentricity-inclination coupling not present in the Laplace-Lagrange eigenvalue solution. 

\begin{figure*}[t]
\includegraphics[width=\textwidth]{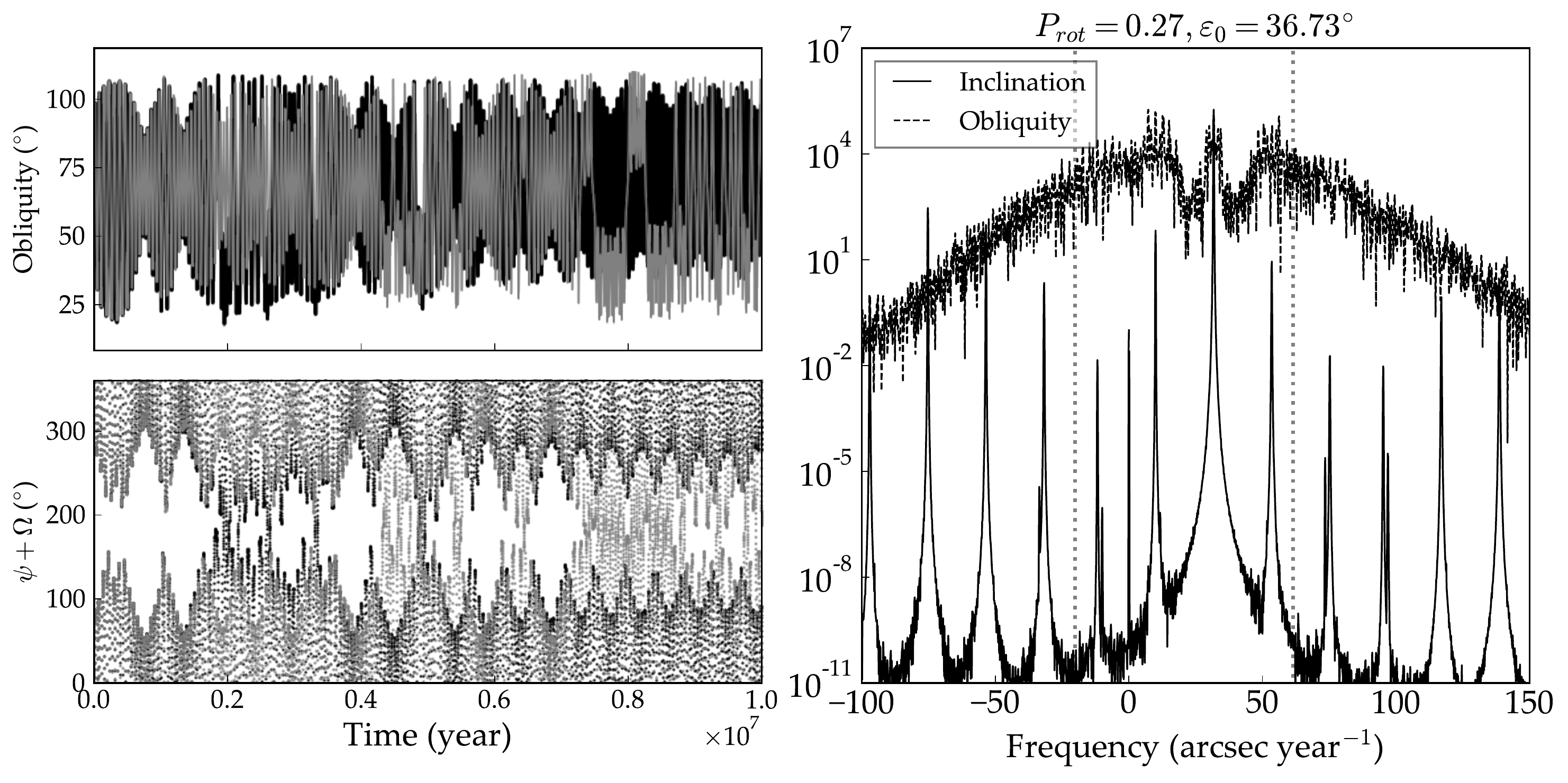}
\caption{\label{sys1chaos} Evolution of the obliquity and the angle $\psi + \Omega$ (left) for the Earth-mass planet a point {\bf h} in Figure \ref{sys1oblvsrot}, and the power spectra of the inclination and obliquity variables (right). Vertical dashed lines represent the minimum and maximum axial precession rates, which vary with obliquity. The broad peak of the obliquity variables is indicative of chaotic motion.}
\end{figure*}

In regions where the obliquity amplitude is $\gtrsim 100^{\circ}$, the motion becomes chaotic---the obliquity undergoes irregular motion and its power spectrum displays a single, extremely broad peak. Figure \ref{sys1chaos} shows this for point {\bf h}. 
The left-hand panels show the obliquity and $\psi+\Omega$ evolution for 10 Myrs in black. The gray curves are the result of slightly different initial conditions---specifically, $\varepsilon_0 = 36.72^{\circ}$, a difference of $0.01^{\circ}$. The conditions in the two simulations diverge at $\sim 2$ Myrs. 
The most probable explanation for this chaotic motion is resonance overlap \citep{chirikov1979,wisdom1980}, in which multiple resonances become active and lead to irregular, unpredictable motion. Though a rigorous calculation of the oscillation zones of the active resonances is outside the scope of this work, we can gather some information from the power spectra in Figure \ref{sys1chaos}. In the region where $P_{rot} \lesssim 0.5$ days and $\varepsilon_0 \lesssim 60^{\circ}$, the initial precession rate of the planet's spin axis is between $\sim 30$ arcsec year$^{-1}$ and $\sim 60$ arcsec year$^{-1}$. This places the planet amidst two strong inclination frequencies at $\sim 31$ arcsec year$^{-1}$ and $\sim 55$ arcsec year$^{-1}$. Because the precession rate varies strongly with $\varepsilon$ in this region and the obliquity oscillations are so large, the spin axis passes through both resonances easily---referring to the right panel of Figure \ref{sys1chaos}, we can see that the precession rate (vertical dashed lines) crosses both, as well as the slower frequency at $\sim 9$ arcsec year$^{-1}$. 

We simulated the same parameter space as in the right panel of Figure \ref{sys1oblvsrot}, but with the initial inclination reduced in $5^{\circ}$ increments. At inclinations at and below $10^{\circ}$, the expected $\cos{\varepsilon}$ shape of the secular resonances is recovered, and the obliquity oscillations are reduced. In the case we show here, where the mutual inclination is increased to $\sim 20^{\circ}$, the obliquity oscillations become larger and the inclination frequency peaks broaden, so that the precession rate, $R(\varepsilon)$, passes through or near to several strong inclination peaks. This could explain the complex structure of Figure \ref{sys1oblvsrot}, the multiple frequencies in the obliquity evolution and intermittant circulation of $\psi+\Omega$ in Figure \ref{orb2sys1}, and the chaotic motion of $\varepsilon$ in some regions. 

We also simulated this parameter space with $\psi_0$ rotated by $180^{\circ}$. It is noteworthy that in that case, the islands at points {\bf e}, {\bf f}, and {\bf g} do \emph{not} appear, indicating the importance of the initial orientation of the spin axis. This adds further support to the hypothesis that these regions represent Cassini-like states resulting from mutual inclinations in the 4th order theory.

\section{DISCUSSION}
\label{sec:discuss}
In our simulations of Kepler-62 and TSYS, we see that the amplitude of the obliquity variations is largest in secular spin-orbit resonances, in which the natural precession frequency of the planet's spin axis is near to the frequency of inclination variations. The planet's rotation rate and initial obliquity play an extremely important role, especially in planar systems like Kepler-62. In our TSYS, in which eccentricity and mutual inclination are allowed be somewhat large, we see the appearance of higher order secular resonances that depend strongly on the initial $e$ and $i$.

For the coplanar system Kepler-62, secular spin-orbit resonances are the largest driver of obliquity evolution of planet f. These resonances are easily identifiable via the Laplace-Lagrange theory and depend intimately on the rotation rate and initial obliquity. While there are a number of studies that lay out possible means of determining rotation rate and obliquity for exoplanets \citep{cowan2009,fujii2012,cowan2013,schwartz2016}, we are probably a decade or more away for doing such for small mass planets. In the mean time, modeling studies such as this may be helpful in predicting possible climate states for planet such as this, which may then be verified by visible phase curve observations or spectroscopic detections.

Cassini states are possible for a planet like Kepler-62 f. In the vicinity of Cassini states, the planet's obliquity oscillates with large amplitude, however, if a dissipational force (like tidal friction) is at work, the planet could be driven into one of the stable states. State 2 is particularly interesting because it is stable and it can occur at high obliquity, provided the rotation rate is rapid enough. Kepler-62  f might exist in a kind of sweet spot, where the tides from the host star are not strong enough to synchronize its rotation rate, but might be strong enough to push it into a Cassini state. It may be worth testing this by coupling the orbital and obliquity models to a tidal model.

While certainly a powerful influence on climate, it is still not clear how likely terrestrial (and possibly habitable) planets are to exist in mutually inclined systems. The triple Jovian system $\upsilon$ Andromedae certainly shows that strongly inclined systems exist, at least for high mass planets \citep{mca2010}. Transiting surveys such as \emph{Kepler} have shown the prevalence of compact, coplanar super-Earth systems in the galaxy, however, with this detection method we may be missing many inclined systems. The only truly unbiased way of detecting mutually inclined systems is to use a combination of techniques, such as radial velocity and astrometry, or high-resolution spectrometry to detect molecular lines associated with planets. Thus far, these detection methods cannot reliably probe down to low mass, terrestrial bodies. 

On the theory end, numerous studies have shown that instabilities resulting in the loss of one or more planets from a system can result in high inclinations and high eccentricities in the remaining planets \citep{ford2005,raymond2009,barnes2011}. Even our own solar system may have had such an event in its early history \citep{gomes2005,batygin2012}. The additional complications associated with binary stars and galactic effects \citep{kaib2013} add further weight to the idea that many systems will have experienced such scattering events.

Though, for simplicity, we model planets without large moons (like Earth's) here, a few words are in order. It is commonly understood that in the absence of our moon, Earth would undergo much large obliquity oscillations than it does \citep{laskar1993,lissauer2012,li2014}. Before we assume that planets which have large moons are safe from large variations, we should consider how it is that Earth's moon actually stabilizes its obliquity. The moon, because of its mass and proximity, induces a torque on Earth's equatorial bulge that is stronger than the Sun's, and this torque significantly increases the planet's axial precession rate. This speed up protects Earth from large obliquity swings, because it is far from the secular frequencies of the solar system, and hence secular spin-orbit resonances are avoided. When we consider the tidal effect of the moon, that is, the slowing of Earth's rotation rate, the matter becomes more complicated. This is because the axial precession rate is a also strong function of rotation rate---in fact, it is predicted that the moon will eventually drag our planet into a secular spin-orbit resonance \citep{ward1982}. Furthermore, Earth's rotation rate might be very different without the moon and its tidal tug. Thus, we argue that we should not simply assume that a large moon is beneficial (or detrimental, for that matter) to climate and habitability.  Exoplanets with large moons may be driven into and out of secular spin-orbit resonances as their moons alter the planetary rotation rate.

\begin{figure}[t]
\includegraphics[width=0.5\textwidth]{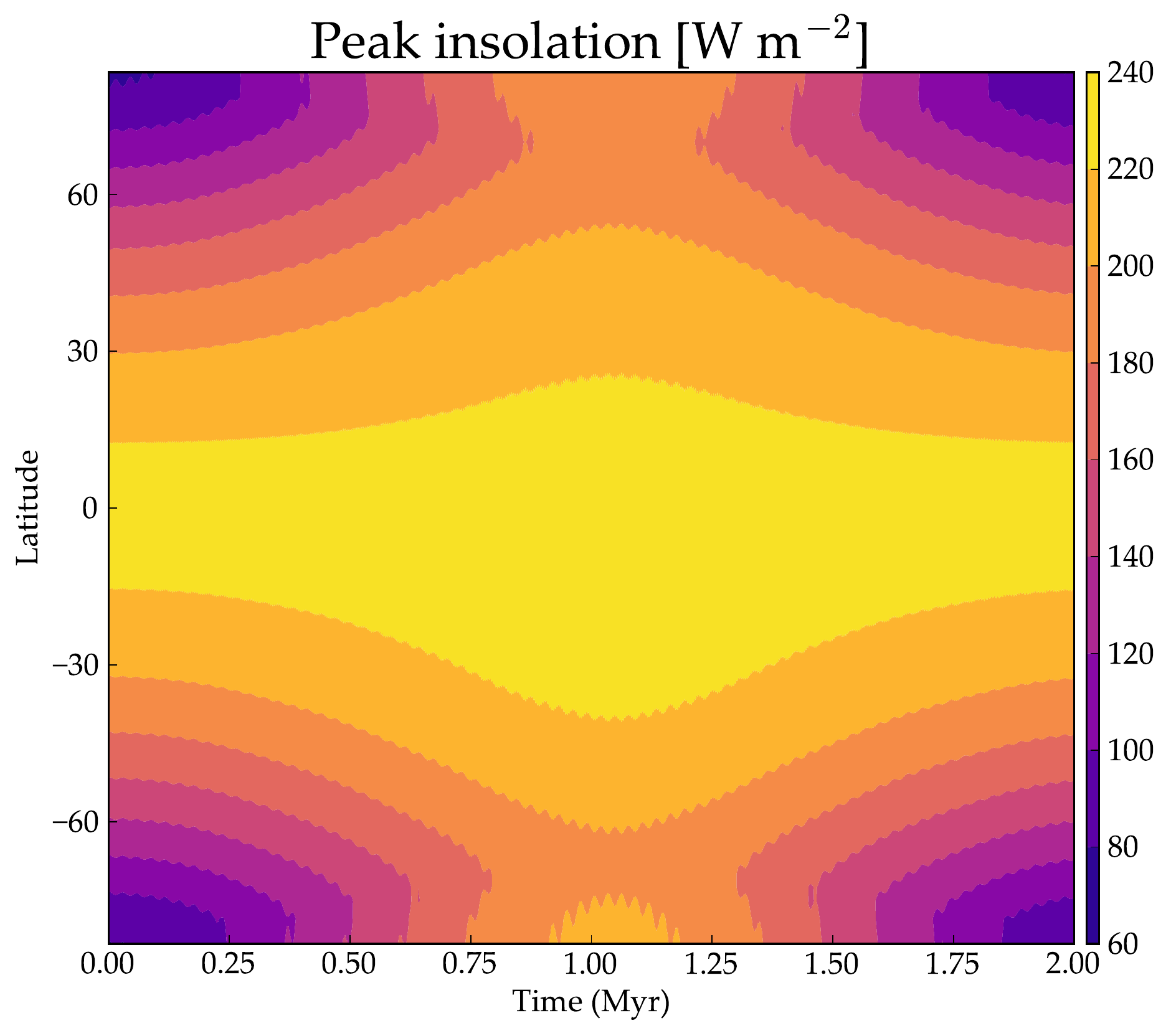}
\includegraphics[width=0.5\textwidth]{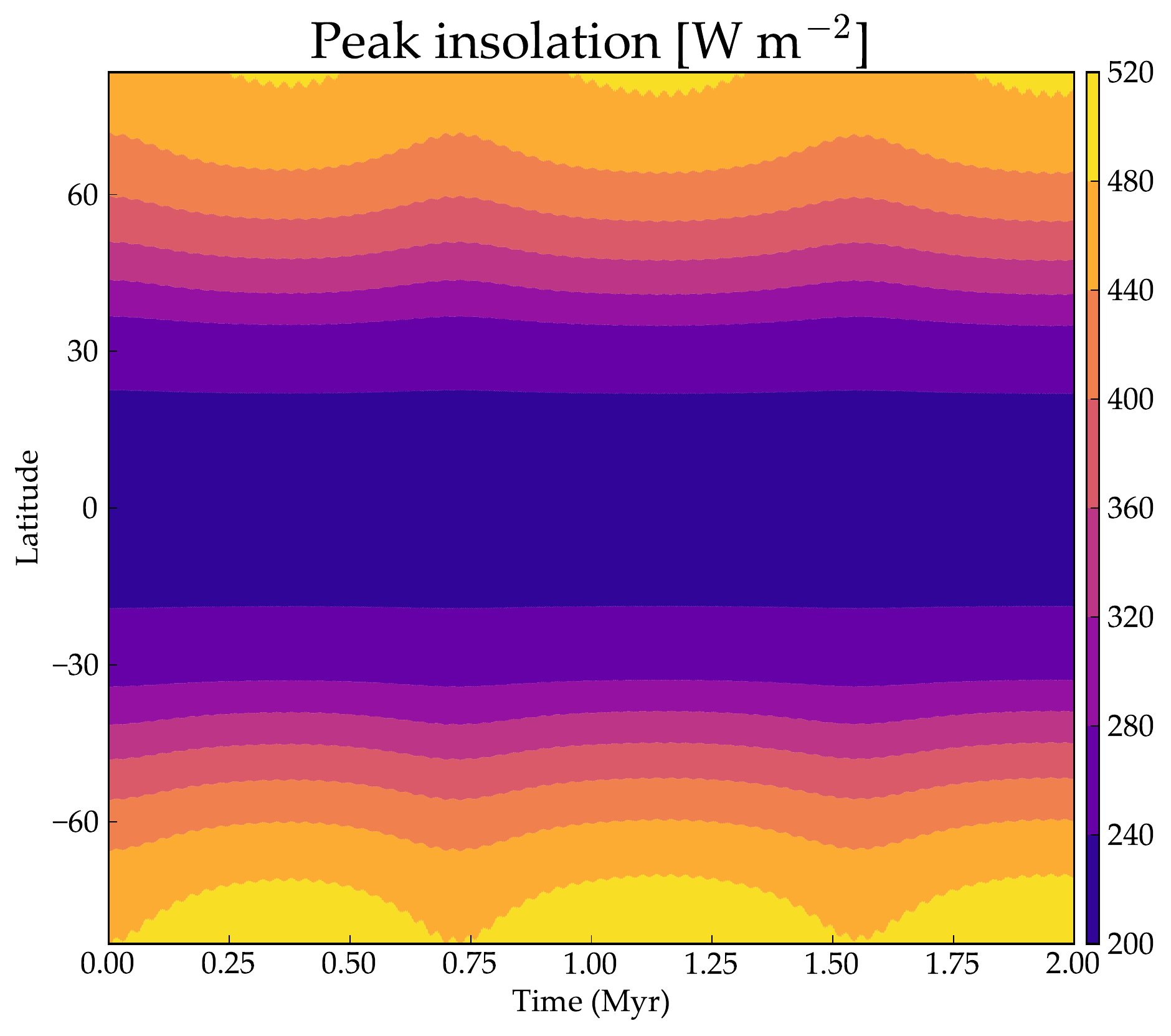}
\caption{\label{k62insol} Peak insolation curves over time for Kepler-62 f as a function of latitude, for two secular resonance cases. The left panel is point {\bf b} at $P_{rot} = 1.09$ days and $\varepsilon_0 = 7.5^{\circ}$, the right at $P_{rot}=0.6$ days and $\varepsilon_0 = 52.5^{\circ}$.}
\end{figure}

We expect secular resonances of this nature will have a substantial impact on the climate, if the atmosphere is at all similar to Earth's. Figure \ref{k62insol} shows the insolation (stellar flux) received by the Kepler-62 f in the secular resonance (point {\bf b} in Figure \ref{k62oblmap} and a case at $P_{rot}=0.6$ days and $\varepsilon_0 = 52.5^{\circ}$), for all latitudes. The insolation shown here is the peak summertime value at each latitude, calculated from the formulae in \cite{berger1978}. For this calculation, the stellar constant is $\sim 573.18$ W m$^{-2}$, about $42\%$ of Earth's value. The annual insolation changes dramatically in the low starting obliquity case, especially for the poles, as the obliquity oscillates. If a planet like this has surface water, we might expect it to have severe ice age cycles. In the high obliquity case, the poles receive more intense seasonal insolation than the equator. The obliquity oscillation is much smaller than the low obliquity case, but the poles still undergo variations in peak stellar flux of $\approx 40$ W m$^{-2}$. Our following paper will address how this impacts the surface temperature and overall habitability of the planet.

How might we determine if an exoplanet is undergoing Milankovitch cycles, given that we cannot observe planets for tens of thousands of years and are unable to travel there to collect rock records? It is worth mentioning that Milankovitch cycles have been inferred for the planet Mars from images of the polar regions taken from orbit \citep{laskar2002}. Though we are technologically far away from the possibility of doing something similar for small HZ planets, several studies have already begun to work out techniques for creating 2-D maps of a planetary surface \citep{fujii2012,cowan2013,kawahara2016,schwartz2016}. This technique will likely require a high-contrast, large aperture telescope, like the recently studied \emph{Large UltraViolet Optical and Infrared surveyor (LUVOIR)} mission \citep{bolcar2016}, capable of directly imaging terrestrial planets. As shown in \cite{schwartz2016}, the technique is capable of determining longitudinal brightness and color variations, obliquity, and, in some cases, latitudinal surface variations. From this, we may be able to determine some limited history of the planet's climate, if, say, the planet has large ice features at latitudes inconsistent with its current obliquity. Kepler-62 f, if it has an obliquity oscillation similar to what we find here, would be a difficult test case because the obliquity evolves rather slowly, which would allow the ice latitude to equilibrate more easily. It would be much easier to observe the inconsistency for a terrestrial planet with a giant companion, like our TSYS, because the obliquity oscillation is more rapid.

From orbital information, it is certainly reasonable to determine whether Milankovitch cycles are a possibility. If the orbital/obliquity variations are large enough to trigger a climate catastrophy (\emph{e.g}, a snowball state), that might be observable through the planet's albedo---this is the subject of our followup study. If the rotation rate and obliquity can be constrained through surface mapping, as discussed above, the case for or against Milankovitch cycles can be further strengthened.

Milankovitch cycles may influence the climates of planets with surface oceans, but would probably have little or no effect on subsurface oceans. Hence, the habitability of planets analogous to Europa or Enceladus would not be affected by this kind of dynamical evolution, unless tidal heating of the interior is significantly modified by obliquity or eccentricity variations. However, the interaction of life with the atmosphere will likely provide the only biosignatures detectable across interstellar distances, hence our focus on surface habitability.

\section{CONCLUSIONS}
\label{sec:conclu}

We have presented a secular model for the orbital and obliquity oscillations associated with planetary perturbations. The speed of the model allows us to explore broad regions of parameter space and to rapidly identify interesting dynamical phenomena. We apply this model to several test cases of Earth-like (or potentially Earth-like) exoplanets to understand how climate and habitability might be affected by planetary companions. 

 We studied three systems here, with a focus on secular spin-orbit resonances and Cassini states. Secular resonances can potentially have a dramatic impact on climate by inducing strong obliquity variations. In the following paper, we explore whether such obliquity variations, combined with eccentricity variations and precession, make planets more or less susceptible toward climate states that might be render the surface inhospitable to life. 

We must emphasize that planetary systems are extremely complex, and that habitable zones based simply on semi-major axis may be fundamentally limited. In assessing a planet's potential to host life, it is important that companion planets be looked for and considered as possible complications, especially for planets that may have not undergone strong tidal evolution.   

This study demonstrates the importance of companion planets and orbital architecture when assessing the habitability of exoplanets. Used in tandem with climate models, this type of analysis will provide testable predictions of atmospheric states for upcoming exoplanet characterization missions, and will aid in the interpretation of potential biosignatures.  In our following paper, we apply a simplified climate model to further explore this problem.
\clearpage

\acknowledgements This work was supported by the NASA Astrobiology Institute's Virtual Planetary Laboratory under Cooperative Agreement number NNA13AA93A. This work was facilitated though the use of advanced computational, storage, and networking infrastructure provided by the Hyak supercomputer system at the University of Washington. The results reported herein benefited from the authors' affiliation with the NASA's Nexus for Exoplanet System Science (NExSS) research coordination network sponsored by NASA's Science Mission Directorate.

\bibliographystyle{aasjournal}
\bibliography{paper1_edit2}
\appendix
\section{The disturbing function and its derivatives}
\label{disturbing}

Here we present, for the sake of completeness, the disturbing function as used in \texttt{DISTORB}, in the variables $h,k,p,$ and $q$ (see Section \ref{distorb}). These were originally derived by \cite{ellis2000}; we have simply applied coordinate transformations and calculated derivatives with respect to the new coordinates. This disturbing function, in its original form, can also be seen in \cite{md1999}. We will not restate the semi-major axis functions, $s_1, s_2, s_3$ and so on, in this paper, as they are taken directly from {bf Table B.3} of \cite{md1999}.

The secular disturbing function, for any pair of planets, is:
\begin{equation}
\mathcal{R} = \frac{\mu'}{a'}\mathcal{R}_D,
\end{equation}
for the inner body and:
\begin{equation}
\mathcal{R'} = \frac{\mu}{a'}\mathcal{R}_D,
\end{equation}
for the outer body. Here, $a'$ is the semi-major axis of the exterior planet and the mass factors are $\mu = \kappa^2 m$ and $\mu' = \kappa^2 m'$, where $m$ is the mass of the interior planet and $m'$ is the mass of the exterior planet. Finally,
\begin{equation}
\mathcal{R}_D = \text{D}0.1 + \text{D}0.2 +\text{D}0.3 +\dots,
\end{equation}
where the terms D0.1, D0.2, and so on, are given in Table \ref{distfxn}.

\begin{table}[h]
\caption{\textbf{Disturbing function}}
\begin{tabular}{lp{0.4\textwidth}}
\hline\hline \\ [-1.5ex]
Term &   \\ [0.5ex]
\hline \\ [-1.5ex]
D0.1 & $f_1 + (h^2+k^2+h'^2+k'^2)f_2+(p^2+q^2+p'^2+q'^2)f_3$\\
	&~$+(h^2+k^2)^2 f_4+(h^2+k^2)(h'^2+k'^2)f_5+(h'^2+k'^2)^2 f_6$\\
	&~$+[(h^2+k^2)(p^2+q^2) + (h'^2+k'^2)(p^2+q^2)$\\
	&~~$ + (h^2+k^2)(p'^2+q'^2)+(h'^2+k'^2)(p'^2+q'^2)] f_7$\\
	&~$ +[(p^2+q^2)^2+(p'^2+q'^2)^2] f_8 + 
(p^2+q^2)(p'^2+q'^2)f_9$   \\[0.5ex]
\hline \\[-1.5ex]
D0.2 & $(h h' + k k') [f_{10} + (h^2+k^2)f_{11}+(h'^2+k'^2)f_{12}$\\
  &~~$+(p^2+q^2+p'^2+q'^2)f_{13}]$\\[0.5ex]
\hline \\[-1.5ex]
D0.3 & $(p p' + q q') [f_{14} + (h^2+k^2+h'^2+k'^2)f_{15}$\\
  &~~$+(p^2+q^2+p'^2+q'^2)f_{16}]$\\[0.5ex]
\hline \\[-1.5ex]
D0.4 & $(h^2 h'^2 - k^2 h'^2 - h^2 k'^2 +k^2 k'^2+4 h h' k k')f_{17}$\\[0.5ex]
\hline \\[-1.5ex]
D0.5 & $(h^2 p^2 - h^2 q^2 - k^2 p^2 + k^2 q^2 +4 h k p q)f_{18}$\\[0.5ex]
\hline \\[-1.5ex]
D0.6 & $[h h' (p^2-q^2) - k k' (p^2-q^2) + 2 p q (h k'+k h')]f_{19}$\\[0.5ex]
\hline \\[-1.5ex]
D0.7 & $(h'^2 p^2 - h'^2 q^2 - k'^2 p^2 + k'^2 q^2 +4 h' k' p q)f_{20}$\\[0.5ex]
\hline \\[-1.5ex]
D0.8 & $(h^2 p p' - h^2 q q' - k^2 p p' + k^2 q q' +2 h k p' q + 2 h k p q')f_{21}$\\[0.5ex]
\hline \\[-1.5ex]
D0.9 & $[(h h' +k k') (p p' + q q') +(h k' -k h') (p q' - q p') ]f_{22}$\\[0.5ex]
\hline \\[-1.5ex]
D0.10 & $[(h h' +k k') (p p' + q q') +(h k' -k h') (q p' - p q') ]f_{23}$\\[0.5ex]
\hline \\[-1.5ex]
D0.11 & $[(h h' - k k') (p p' - q q') +(h k' +k h') (p q' +  q p') ]f_{24}$\\[0.5ex]
\hline \\[-1.5ex]
D0.12 & $(h'^2 p p' - h'^2 q q' - k'^2 p p' + k'^2 q q' +2 h' k' p' q + 2 h' k' p q')f_{25}$\\[0.5ex]
\hline \\[-1.5ex]
D0.13 & $(h^2 p'^2 - h^2 q'^2 - k^2 p'^2 + k^2 q'^2 +4 h k p' q')f_{18}$\\[0.5ex]
\hline \\[-1.5ex]
D0.14 & $[h h' (p'^2-q'^2) - k k' (p'^2-q'^2) + 2 p' q' (h k'+k h')]f_{19}$\\[0.5ex]
\hline \\[-1.5ex]
D0.15 & $(h'^2 p'^2 - h'^2 q'^2 - k'^2 p'^2 + k'^2 q'^2 +4 h' k' p' q')f_{20}$\\[0.5ex]
\hline \\[-1.5ex]
D0.16 & $(p^2 p'^2 - p^2 q'^2 - q^2 p'^2 + q^2 q'^2 +4 p q p' q')f_{26}$\\[0.5ex]
\hline \\[-1.5ex]
\end{tabular}
\label{distfxn}
\end{table}

\end{document}